\documentstyle[epsfig]{aipproc}

\def\beq{\begin{equation}}
\def\eeq{\end{equation}}
\def\ni{\noindent}
\newcommand{\pp}{{_{I\hspace{-0.2em}P}}}
\newcommand{\rr}{{_{I\hspace{-0.2em}R}}}
\input{epsf.sty}

\begin{document}
\title{Deep Inelastic Scattering, Diffraction and all that\thanks{Partially
 supported by CONICET and Agencia Nacional de Promoci\'on Cient\'{\i}fica, Argentina.
} }
\author{C.A. Garc\'{\i}a Canal$^*$, R. Sassot$^{\dagger}$}

\address{$^*$ Laboratorio de F\'{\i}sica Te\'{o}rica\\ Departamento de
F\'{\i}sica, Universidad Nacional de La Plata\\ C.C. 67 - 1900 La
Plata,  Argentina\\ $^{\dagger}$ Departamento de F\'{\i}sica,
Universidad de Buenos Aires\\ Ciudad Universitaria, Pab.1 (1428)
Buenos Aires, Argentina }

\maketitle

\begin{abstract}
These lectures include an introduction to the partonic description
of the proton, the photon and the `colour singlet', as seen in
inclusive and semi-inclusive DIS, in $e^+e^-$ collisions, and in
diffractive processes, respectively. Their formal treatment using
structure, fragmentation, and fracture functions is outlined
giving an insight into the perturbative QCD framework for these
functions. Examples and comparisons with experimental data from
LEP, HERA, and Tevatron are also covered.
\end{abstract}

\section*{Introduction}

The discovery of asymptotic freedom, one of the most significant
properties of strong interactions embodied by Quantum
Chromodynamics (QCD) opened, more than 25 years ago, a new chapter
in our understanding of the structure of matter which has has been
actively followed by theoreticians and experimentalists ever
since. The short distance  structure of hadrons, together with the
production of jets in hadronic collisions are paradigmatic among
the strong interaction phenomena successfully accounted by QCD and
even though the standing of the theory is today well established,
further theoretical refinements and the corresponding experimental
validation renew constantly the original enthusiasm of the high
energy physics community.

These lectures intend to provide an overview of the more recent
topics of high energy collisions related in a  way or another to
perturbative QCD. First we  briefly remind the essentials of QCD,
including the
main features of partons (the quarks and gluons), which are the
true protagonists in the story. Then, we refer to what is known
about the partonic structure of three of the main benchmarks of
QCD, the proton, the photon, and the singlet colour or 'pomeron'.
Finally we will try to draw the connections between their
corresponding structures which in some way relate the physics made in
the three main HEP laboratories.

As usual, many interesting and highly active topics have been
excluded from the lectures in favor of a more detailed analysis of
the covered points. These include, for example, those related to
the spin structure of the proton, which have driven an ongoing
series of polarized experiments and a great deal of theoretical
discussions \cite{Lampe}; heavy flavours, which involve very
subtle theoretical approaches, and perturbative QCD
beyond NLO, which is relevant for the most recent high precision
experiments \cite{BASS}.

\section*{QCD}

The strong interactions among quarks and gluons are described by
Quantum Chromodynamics (QCD), the non-abelian gauge theory based
on the gauge group $SU(3)_C$. Each quark flavour corresponds to a
colour triplet in the fundamental representation of $SU(3)$ and
the gauge fields needed to maintain the gauge symmetry, the
gluons, are in the adjoint representation of dimension 8. Gauge
invariance ensures that gluons are massless. The QCD
Lagrangian may be written as

\beq {\cal{L}}_{QCD} = - \frac{1}{4}\,F^a_{\mu\nu}\,F^{\mu\nu}_a +
\bar{\psi}_i\,(i\,\gamma^{\mu}\,D_{\mu} - m)\,\psi_{i} \eeq where

\beq F^a_{\mu\nu} = \partial_{\mu}G^a_{\nu} -
\partial_{\nu}G^a_{\mu} + g\,f^{abc}\,G_{b\mu}\,G_{c\nu} \eeq
stands for the gluon field tensor, $\psi_i$ are the quark fields
and the covariant derivative is defined by
\[
D_{\mu} = \partial_{\mu} - i\,g\,T_a\,G^a_{\mu}
\]
The strong coupling is represented by $g$ and indices are summed
over $a = 1,...,8$ and over $i=1,2,3$. Finally, $T_a =
\lambda_a/2$ and $f_{abc}$ are the $SU(3)$ generators and
structure constants, respectively, which are related by $
[T_a,T_b] = i\,f_{abc}\,T^c$.

Like in Quantum Electrodynamics (QED), the procedure employed to deal
consistently with the divergences that occur in the computation of
strong interactions beyond the tree level, shows that the actual
strength of the QCD coupling depends on the energy scale of the process. But in
opposition to QED, this renormalized strong coupling is small at
high energy (momentum), going to zero logarithmically, i.e. QCD has the
property of {\it asymptotic freedom}. Consequently, in this regime
perturbation theory is valid and tests against experimental data
can be performed in terms of hadrons. Figure \ref{fig:PQCD}
summarize the basic QCD perturbative processes appearing in
different circumstances.

\setlength{\unitlength}{1.mm} 
\begin{figure}[hbt]
\begin{picture}(150,50)(0,0)
\put(44,-100){\mbox{\epsfxsize12.5cm\epsffile{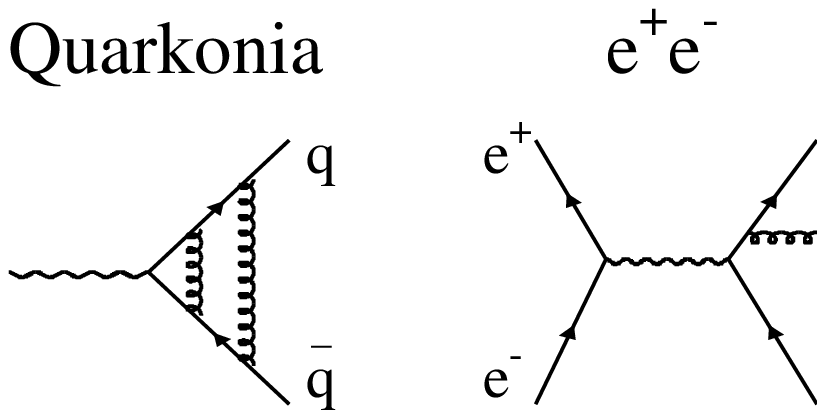}}}
\put(-27,-102){\mbox{\epsfxsize12.5cm\epsffile{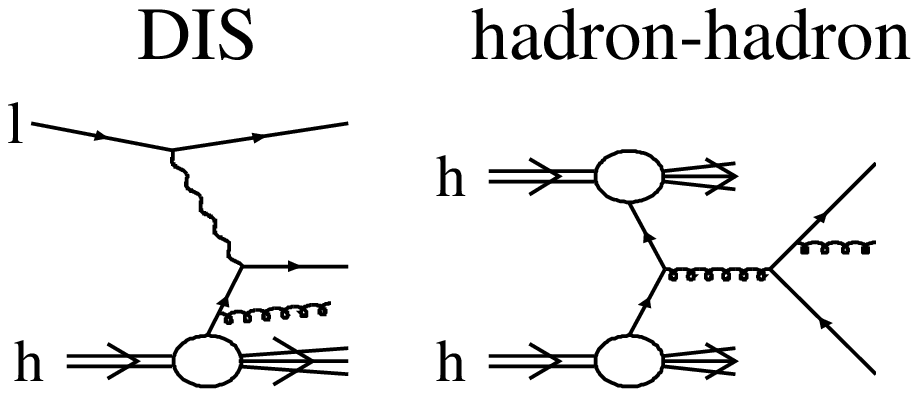}}}
\vspace{10pt}
\caption{Basic processes in perturbative QCD.}
\label{fig:PQCD}
\end{picture}
\end{figure}

Experiments with $e^-\,e^+$ colliders provide clean results for
QCD tests. Recently, a huge amount of experimental data came from
the HERA electron-proton collider \cite{revHE} and also from the
Tevatron at Fermilab \cite{revFE}. In both cases, there is a
hadronic remnant that make the analysis a little more involved.
All this experimental evidence support the existence of quarks
being colour triplets of spin $1/2$ and of gluons being vector
octets. Moreover, the presence of the QCD coupling has manifested
itself in different measurements, as well as the above mentioned
property of asymptotic freedom. This information comes mainly
from the study of the so called two- and three-jets events \cite{DIS}.

When a given process needs a higher order in perturbation theory
to be known, it is necessary to compute not only the renormalized
strong coupling constant but also the appropriate corrections to
the relevant cross-sections. As is usual in Quantum Field Theory,
a regularization-renormalization procedure is in order, just to
absorb divergences into the definition of physical quantities.
This prescription requires the introduction of a new scale $\mu$,
fixing the renormalization point, and all renormalized quantities
begin to depend on it. Nevertheless, different prescriptions must
end with the same predictions for observables.

In order to illustrate how the general procedure works, ending
with the Renormalization Group equations that guarantee that
physical observables do not depend on the scale $\mu$, let us show
what happens with Green functions. Just to remember the procedure,
let us begin with a single particle irreducible Green function
$\Gamma$. In general, to control divergences, one has to introduce
an ultra-violet cut-off $\Lambda$, or the equivalent dimensional
regularization parameter, in the loop momentum integral defining
the $\Gamma$. In a renormalizable theory, as QCD is, a
renormalized Green function is defined as
\[
\Gamma_R(p_i, g, \mu) =
Z_{\Gamma}(g_0,\Lambda/\mu)\,\Gamma_U(p_i,g_0,\Lambda)
\]
where $p_i$ stands for the external particle momenta, $g_0$ and
$g$ are the bare and the renormalized couplings, respectively.
This $\Gamma_R$ is then finite in the limit $\Lambda \rightarrow
\infty$ but it depends on the scale at which the value of the
renormalized quantities are fixed, the prescription parameter
$\mu$. The function $Z_{\Gamma}$ is a product of renormalization
factors. Due to the fact that the unrenormalized $\Gamma_U$ is
obviously independent of $\mu$, one has to demand
\[
\frac{d \Gamma_U}{d \mu} = 0
\]
and consequently, the Renormalization Group Equation (RGE)
 \beq
\left( \mu\,\frac{\partial}{\partial \mu} +
\beta\,\frac{\partial}{\partial g} + \gamma \right)\,
\Gamma_R(p_i, g, \mu) = 0 \eeq has to be verified. Here $\gamma$
is the {\it anomalous dimension}, depending on the particular
Green function under consideration, and the  {\it beta-function}
is universal
 \beq
 \gamma =
\frac{\mu}{Z_{\Gamma}}\,\frac{\partial Z_{\Gamma}}{\partial \mu}
\,\,\,\,\,\,\,\,\,\,\,\,\,\,
\beta(g) = \mu\,\frac{\partial g}{\partial \mu}
\eeq

If there is only one large momentum scale $Q$, or $Q^2$ as it is
standard to quote, as it is the case here, one can express all
$p_i$ in terms of a fixed fraction $x_i$ of $Q$. Then, defining
the so called evolution variable

\beq t = \frac{1}{2}\,\ln \left(\frac{Q^2}{\mu^2} \right)
 \eeq
 it
is possible to introduce the momentum dependent, or {\it running}
coupling through the integral
\[
t = \int^{g(t)}_{g(0)} \frac{d g^{\prime}}{\beta(g^{\prime})}
\]
and the general solution of the RGE reads
\[
\Gamma(t,g(0),x_i) = \Gamma(0,g(t),x_i)\,\, exp \left[
\int^{g(t)}_{g(0)} dg^{\prime} \,\frac{\gamma(g^{\prime})}
{\beta(g^{\prime})} \right]
\]
This solution explicitly shows that the $Q$-scale dependence of
$\Gamma$ arises entirely through the running coupling $g(t)$.
Introducing now the usual notation \beq \alpha_s =
\frac{g^2}{4\pi}
 \eeq
 one can expand the beta-function in a power
series in $\alpha_s$ \beq \beta(\alpha_s) = \mu\,\frac{\partial
\alpha_s}{\partial \mu} = - \frac{\beta_0}{2 \pi}\,\alpha^2_s -
\frac{\beta_1}{4 \pi^2}\,\alpha^3_s - ... \eeq
 where it results \cite{beta} that
\begin{eqnarray}
\beta_0  =  11 - \frac{2}{3}\,N_f  \,\,\,\,\, \,\,\,\,\, \beta_1  =  51 -
\frac{19}{3}\,N_f
\end{eqnarray}
Here $N_f$ indicates the number of flavours that can be excited
(with mass less than $\mu$) at the scale $\mu$.

It is clear that the solution of the differential equation for
$\alpha_s$ introduces a constant, called $\Lambda_{QCD}$, which
has to be fixed by using experimental data. The resulting
$\alpha_s$ can be written as
\begin{equation} \label{eq:al}
\alpha_s(\mu, \Lambda_{QCD})  =
\frac{4\,\pi}{\beta_0\,\ln(\mu^2/\Lambda_{QCD}^2)}\, \left\{ 1 -
\frac{2\,\beta_1}{\beta_0^2}\,\frac{\ln[\ln(\mu^2/\Lambda_{QCD}^2)]}
{\ln(\mu^2/\Lambda_{QCD}^2)}\right\}
\end{equation}

\begin{figure}[t!] 
\centerline{\epsfig{file=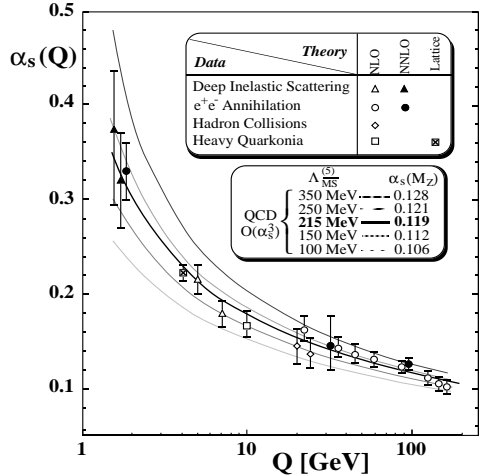,height=2.5in,width=2.5in}}
\vspace{10pt}
\caption{QCD running coupling.}
\label{fig:ALFA}
\end{figure}

\ni This expression for the running coupling shows clearly the
property of asymptotic freedom of QCD, i.e., the coupling vanishes
when the scale becomes asymptotic, namely $\mu \rightarrow
\infty$. Consequently, in this momentum regime, perturbation
theory is valid.

A very clear quantitative test of perturbative QCD is provided by
the measurement of $\alpha_s$ in different processes at different
scales $Q^2$. In Figure \ref{fig:ALFA} there is a summary of the
various determinations of $\alpha_s$ \cite{9810316}.

The present world average \cite{PDG} for the coupling at the $Z^0$
mass is
\[
\alpha_s(M_Z) = 0.119 \pm 0.002
\]
which implies
\[
\Lambda_{QCD}^{\overline{MS}} = 220 +78-63\, MeV
\]
corresponding to five flavours excited and in the conventional
 $\overline{MS}$ prescription commonly used.

\section*{Proton}

Having reviewed the essentials of QCD, we now proceed with the
analysis of one of the most rich benchmarks of perturbative QCD
which is the short distance structure of the proton.

We start this section giving a set of definitions of the
commonly used relativistic invariants related to Deep Inelastic
Scattering, DIS, one of the fundamental experimental tools for
hadron analysis, together with the corresponding formulae for the
neutral and charged current cross-sections, where the structure
functions are introduced. These functions will be latter expressed
in terms of the quark-parton model including QCD corrections. For
a more detailed treatment see \cite{DIS}

\subsection*{Lorentz Invariants}

\setlength{\unitlength}{1.mm} 
\begin{figure}[b!]
\begin{picture}(150,60)(0,0)
\put(30,-72){\mbox{\epsfxsize10.5cm\epsffile{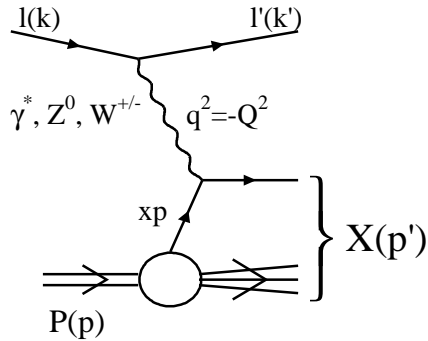}}}
\caption{DIS kinematics.}
\label{fig:DIS}
\end{picture}
\end{figure}

The scattering of a lepton from a proton (or in general a nucleon,
a hadron or a nucleus), at high enough $Q^2$, can be viewed as the
(elastic) scattering of the lepton from a quark or antiquark
inside the proton mediated by the exchange of the corresponding
virtual vector boson $\gamma^{\star}, \, W^{\star}$ or
$Z^{\star}$. Consequently, when the process is totally inclusive
(one integrates over all final hadronic activity), can be fully
described by two Lorentz invariants. With the momentum assignment
of Figure \ref{fig:DIS} one can construct the following
invariants:
\beq s = (p + k)^2,  \,\,\,\,\,  t = (p - p^{\prime})^2,  \,\,\,\,\, W^2 = (p + q)^2, \,\,\,\,\, Q^2  = - (k^{\prime} - k)^2  =  - q^2
\eeq
defined as the centre of mass energy squared, the square of the four-momentum transfer  between the proton and the final hadronic state, the invariant mass squared of the final hadronic state, and four momentum transfer squared between the lepton and the proton, respectively. It is also convenient to introduce  dimensionless invariants (scaling variables)
\beq y = \frac{p \cdot q}{p \cdot k} \,\,\,\,\, , \,\,\,\,\,  x = \frac{Q^2}{2\,p \cdot q}\eeq
i.e. the inelasticity of the scattered lepton, and fraction of the proton momentum carried by the struck quark, or Bjorken variable, respectively.


Notice finally that, ignoring masses,
\beq
Q^2 = s\,x\,y \,\,\,\,\,\,\,\,\,\, W^2 = Q^2\,\frac{1 - x}{x}
\eeq
so that $Q^2_{max} = s$, the center of mass energy squared, and
small values of $x$ imply increasing $W$. As we have
already said, in principle one is allowed to use any pair of these
invariants to describe totally inclusive DIS. Usually $x$ and
$Q^2$ are the preferred ones.

Another interesting point to remark concerns the resolving power
of a DIS experiment. Clearly, the size $d$ one can resolve inside
the nucleon becomes smaller for large photon, or gauge boson in
general, virtuality, namely
\[
d \sim \frac{\hbar\,c}{Q} \simeq 0.2\,\frac{GeV\,fm}{Q}
\]
The `magnifying power' is then $d \simeq 10^{-14}\,cm$
for $Q^2 = 4\,GeV^2$, and $d \simeq 10^{-16}\,cm$ and for $Q^2 =
40,000\, GeV^2$.

\subsection*{Experimental Reconstruction}

\begin{figure}[b!]  
\centerline{\epsfig{file=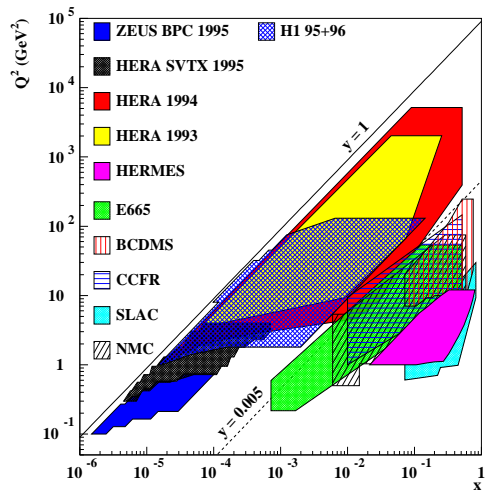,height=2.5in,width=2.5in}}
\vspace{10pt}
\caption{Coverage of the ($x,Q^2$) plane by
the various DIS experiments \label{fig:KDIS} }
\end{figure}

Different DIS experiments cover different regions of the
kinematical plane ($x,Q^2$)
 as shown in Figure \ref{fig:KDIS} \cite{9712505}. It is particularly interesting the very extended
 range covered by HERA: $Q^2 \simeq 0.2\,-\,10^4\, GeV^2$
 and $x \simeq  10^{-5}\,-\,10^{-1}$.

The kinematics of the DIS events, namely the two invariants
required to specify DIS processes, can be determined (particularly
at HERA) from measurements on the electron (the lepton) alone, on
the final hadronic state corresponding to the struck quark alone
or on a mixture of both. In general, the preferred method depends
on the particular kinematical region of interest, strongly
correlated with the detector performance.

\ni {\it The electron method:} The input are the energy $E^{\prime}_e$ of the final electron
and the electron scattering angle $\theta_e$ measured with respect
to the proton beam direction ($\theta_e = 180^\circ$ means zero
electron scattering angle)
\[
y_e = 1 - \frac{E^{\prime}_e}{E_e}\,\sin^2\frac{\theta_e}{2}
\,\,\,\,\,\,\,\,\,\,
Q^2_e= 4\,E_e\,E^{\prime}_e\,\cos^2 \frac{\theta_e}{2}
\]

\ni {\it The hadron method:} The input is the hadronic energy $E_h$ and the three-momentum
components  of the hadronic system. They are computed as the sum
over all final state hadrons $h$
\[
y_h= \frac{\sum_h (E_h - p_{z,h})}{2\,E_e}
\,\,\,\,\,\,\,\,\,\,
Q^2_h = \frac{p^2_{x,h} + p^2_{y,h}}{1 - y_h}
\]

\ni This method is also known with the name Jacquet-Blondel,
because these authors were able to show that the contribution from
hadrons lost in the beam pipe is negligible.

\ni{\it The sigma method:} This method is based on both electron and hadron measurements.
The denominator of $y_h$ is replaced by a sum over all final state
particles including the scattered electron (certainly equal to
$2\,E_e$ due to energy-momentum conservation). The name of the
method comes from the introduction of the variable $\Sigma = \sum_h (E_h - p_{z,h})$
that appears in
\[
y_{\Sigma} = \frac{\Sigma}{\Sigma + E^{\prime}_e\,(1 - \cos
\theta_e)}
\,\,\,\,\,\,\,\,\,\,
Q^2_{\Sigma} =\frac{E^{\prime 2}_e\, \sin^2 \theta_e)}{1 -
y_{\Sigma}}
\]

\ni Notice that the denominator in $y_{\Sigma}$ is twice the true
incident electron energy in the eventual case when the electron
radiates photons that are not detected.

There are other more sophisticated methods, well adapted to
particular situations, like the so-called double angle method or
the PT method used by ZEUS \cite{revHE}.

\subsection*{Cross-sections}

The fundamental measurement in DIS experiments concerns the
totally inclusive cross-section for
\[
\ell(k) + N(p) \rightarrow \ell^{\prime}(k^{\prime}) + X
(p^{\prime})
\]
as a function of the kinematical variables defined above.
 For charged lepton-nucleon scattering mediated by the neutral current,
 the spin averaged cross-section is given in terms of the structure functions
  $F_2$, $F_L$ and $F_3$
\begin{eqnarray} \label{eq:cs}
\frac{d^2\sigma_{\pm}}{dx\,dQ^2} & = &
\frac{4\,\pi\alpha^2}{x\,Q^4} \\ \nonumber &    & \left[ \left(1 -
y +\frac{y^2}{2}\right)\,F_2(x,Q^2) - \frac{y^2}{2}\,F_L(x,Q^2)\mp
\left( y - \frac{y^2}{2} \right)\,F_3(x,Q^2) \right]
\end{eqnarray}
here $\alpha$ is the QED coupling constant. For $Q^2$ values below
the $Z^0$ scale, the parity violating effects related to $F_3$ are
negligible and all the process is due to $\gamma^{\star}$
exchange. Remember also that the longitudinal structure function
$F_L$, a QCD correction important for large $y$, is defined in
terms of the standard $F_1$ and $F_2$, related to the transverse
and longitudinal $\gamma^{\star}\,N$ cross sections respectively,
as
 \beq F_L(x,Q^2) =
F_2(x,Q^2)\,\left(1+\frac{4\,M_N^2\,x^2}{Q^2}\right) -
2\,x\,F_1(x,Q^2)
 \eeq
 and that in the naive quark-parton model,
really valid at extremely high $Q^2$, where quarks are considered
free, massless, having spin $1/2$ and without any $p_T$ developed,
$F_L$ is zero because the Callan-Gross relation
\[
2\,x\,F_1(x) = F_2(x)
\]
is satisfied. Under this assumptions, the so called Bjorken
scaling is fully satisfied, namely, structure functions are only
function of the $x$ variable.

The ratio $R$ is defined by \beq R(x,Q^2) =
\frac{F_L(x,Q^2)}{F_2(x,Q^2) - F_L(x,Q^2)} =
\frac{\sigma_L}{\sigma_T} \eeq which is obviously zero in the
Callan-Gross limit and can be interpreted as the ratio of
cross-sections for the absorption of transversely and
longitudinally polarized virtual photons on nucleon. The
differential cross-section (\ref{eq:cs}) can be rewritten in terms
of $R(x,Q^2)$ as
\begin{eqnarray}
\frac{d^2\sigma}{dx\,dQ^2}  =  \frac{4\,\pi\alpha^2}{x\,Q^4}
\left[1 - y - \frac{M_N^2\,x^2\,y^2}{Q^2} + \frac{y^2}{2}\,
 \frac{1+\frac{4\,M_N^2\,x^2}
  {Q^2}}{1+R(x,Q^2)} \right]\,F_2(x,Q^2)
\end{eqnarray}
where the parity violating contribution was discarded.

It is possible to specify the structure function $F_2$ in terms of
the partons (or better quarks) within the nucleon as follows
 \beq
F_2(x,Q^2) = x\,\sum_f\,e^2_{q_f}\,\left[q_f(x,Q^2) +
\bar{q}_f(x,Q^2)\right]  \label{eq:qqb}
 \eeq
 where the sum runs
over all momentum distributions of quarks of the different
flavours $f$ excited ($x\,q_f(x,Q^2)$) and antiquarks ($x\,\bar{q}_f(x,Q^2)$)
contained in the nucleon and $e_{q_f}$ stands for the the
different parton-photon couplings, namely their electric charges.
Notice that the previous expression (\ref{eq:qqb}) is valid only
in leading order of perturbation theory. In fact, there one has
included the explicit $Q^2$ dependence which is the effect of
implementing first order perturbative QCD. However, there is a
particular renormalization and factorization scheme that can be
used in higher order QCD, the so called DIS-scheme, where one can
retain that expression at any order of perturbation theory. In
other schemes, like the usual $\overline{MS}$ one, the expressions
for the structure functions are more involved, including in
general $g(x,Q^2)$, the gluon density in the nucleon.

Let us now refer for a moment to the inclusive reaction that goes via charged current
 \beq
\nu_{\mu} + N \rightarrow \mu^{-} + X \eeq where as usual we write
$N$ for an isoscalar nucleon. In the QCD improved parton model,
the corresponding cross section reads
 \beq
 \frac{d^2
\sigma}{dx\,dy} = \frac{2\,G^2_F\,M_N\,E_{\nu}}{\pi}\,\left(
\frac{M_W^2}{Q^2+M_W^2}\right)^2\,\left[x\,q(x,Q^2) +
x\,\bar{q}(x,Q^2)\,(1 - y^2) \right] \eeq 
$M_W$ stands for the charged intermediate boson mass.
$G_F$ is the Fermi constant. The quark and antiquark distribution
functions can be written in terms of the corresponding valence and
sea flavour distributions in a proton as
\begin{eqnarray}
2\,q(x,Q^2) & = & u_v(x,Q^2) + d_v(x,Q^2) + u_s(x,Q^2) +
d_s(x,Q^2)\\ \nonumber
 &  & + 2\,s_s(x,Q^2) + 2\,b_s(x,Q^2) \\
2\,\bar{q}(x,Q^2) & = & u_s(x,Q^2) + d_s(x,Q^2) + 2\,c_s(x,Q^2) +
2\, t_s(x,Q^2)
\end{eqnarray}

One can introduce immediately the corresponding structure function
$F_2$ for neutrino scattering. It is of interest to compare this
$F_2^{\nu}$ with the one for charged lepton DIS written above.
Taking into account the sea corrections, the relation between both
$F_2$ functions results \beq \frac{F_2^{\ell}}{F_2^{\nu}} =
\frac{5}{18}\,\left(1 - \frac{3}{5}\,\frac{s+\bar{s} - c -
\bar{c}}{q + \bar{q}} \right) \eeq
This relation compares well
with experimental data \cite{chek}.

\subsection*{QCD evolution equations}

All the functions introduced above include an explicit $Q^2$
dependence. This fact reflects the presence of interactions among
partons. Clearly, the parton model, that implies exact Bjorken
scaling, has to be modified to include these interactions.
Moreover, quarks and gluons are finally confined within hadrons by
the non-Abelian gauge interaction in QCD. Even if QCD is with us
for more than 25 years, it has not yet been possible to calculate
in detail the structure of hadrons starting from quarks, gluons
and their gauge interactions, making compatible the short distance
almost free partons with the long distance confinement. This fact,
related to the running character of the strong interaction
strength that we have discussed above,  is on the origin of our
continuous use of structure (and fragmentation and fracture)
functions in describing DIS. Fortunately enough, some {\it
factorization theorems}, which are valid in QCD, allows one to
split the problem into a perturbative part and a non-perturbative
part. In other words, the cross-section can be expressed as the
folding of the initial parton distribution function $q_{f/N}$, a
non-perturbative input given the density of partons $f$ in the
nucleon, with a lepton-parton cross-section that can be computed
by means of perturbation theory inside QCD.  This fact can be
schematically written as
\[
\sigma_{\ell N} = \sum_f q_{f/N}\,\otimes\,\sigma_{\ell q}
\]
and consequently, it can be sketched as in Figure \ref{fig:FAC}

\setlength{\unitlength}{0.5mm}

\begin{figure}[hbt]  
\begin{picture}(170,110)(0,0)
\put(50,-155){\mbox{\epsfxsize10.0cm\epsffile{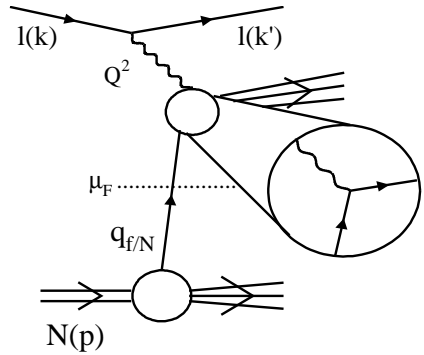}}}
\end{picture}
\caption{Factorization in QCD \label{fig:FAC}}
\end{figure}

It is clear that the non-perturbative part, namely, the part
related to structure
 functions, has to be determined by fitting experimental data. Nevertheless, they are universal in the sense that once obtained from a given particular process, they can be used in connection with any other one.

Regarding QCD as an improvement of the quark-parton model, we can
say that the nucleon is not simply composed of three point-like
quarks, the so called {\it valence} quarks but as soon as $Q^2$
increases, the vector boson ($\gamma^{\star}$, $Z^{\star}$ or
$W^{\star}$) increasingly resolves the composition of the nucleon.
One of the targets that the vector boson could find is one of the quarks,
called {\it sea} quarks, which originate from a gluon, namely
through
\[
g \rightarrow q\,\bar{q}
\]
being this gluon itself radiated from one of the valence quarks.
In other words, increasing $Q^2$ means that the resolution of our
``view" of the nucleon increases, so that the effective number of
partons sharing the nucleon momentum also increases. Consequently, the probability of finding partons with small $x$ has
increased while the corresponding one to large $x$ has decreased.
This process clearly entails a violation of Bjorken scaling
through a $Q^2$ dependence, that was born from QCD interactions.
Consequently, to compare experimental data with QCD predictions.
one has to compute perturbative QCD corrections to the fundamental
process \beq V^{\star} \, q \rightarrow q  \label{eq:Vq} \eeq
Nowadays, these corrections, even if the calculation is very
involved, are well known up to order $\alpha_s^2$ \cite{Van}. We
give here only an introductory discussion of the lowest order
corrections.

The relevant diagrams corresponding to the first terms of the
perturbative expansion in $\alpha_s$ for the process (\ref{eq:Vq})
are presented in Figure \ref{fig:OAS} (We have omitted the vertex
and self energy gluonic corrections, which are associated to the
ultraviolet divergences. These are removed through coupling
constant renormalization, and lead to the running coupling
$\alpha_s(Q^2)$ given in Equation (\ref{eq:al}))

\setlength{\unitlength}{0.5mm}

\begin{figure}[hbt]             
\begin{picture}(170,155)(0,0)
\put(40,90){\mbox{\epsfxsize2.0cm\epsffile{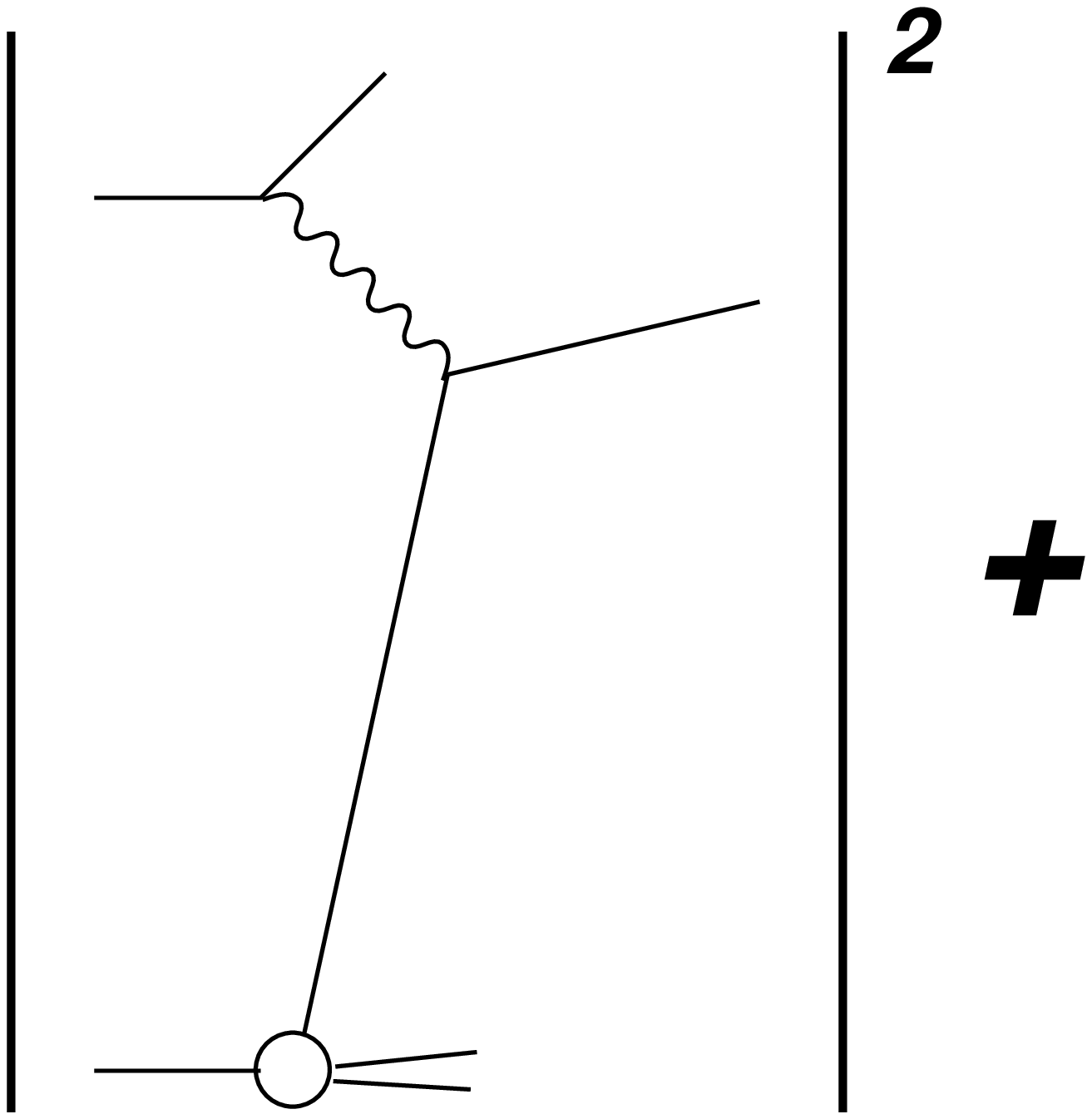}}}
\put(100,90){\mbox{\epsfxsize3.5cm\epsffile{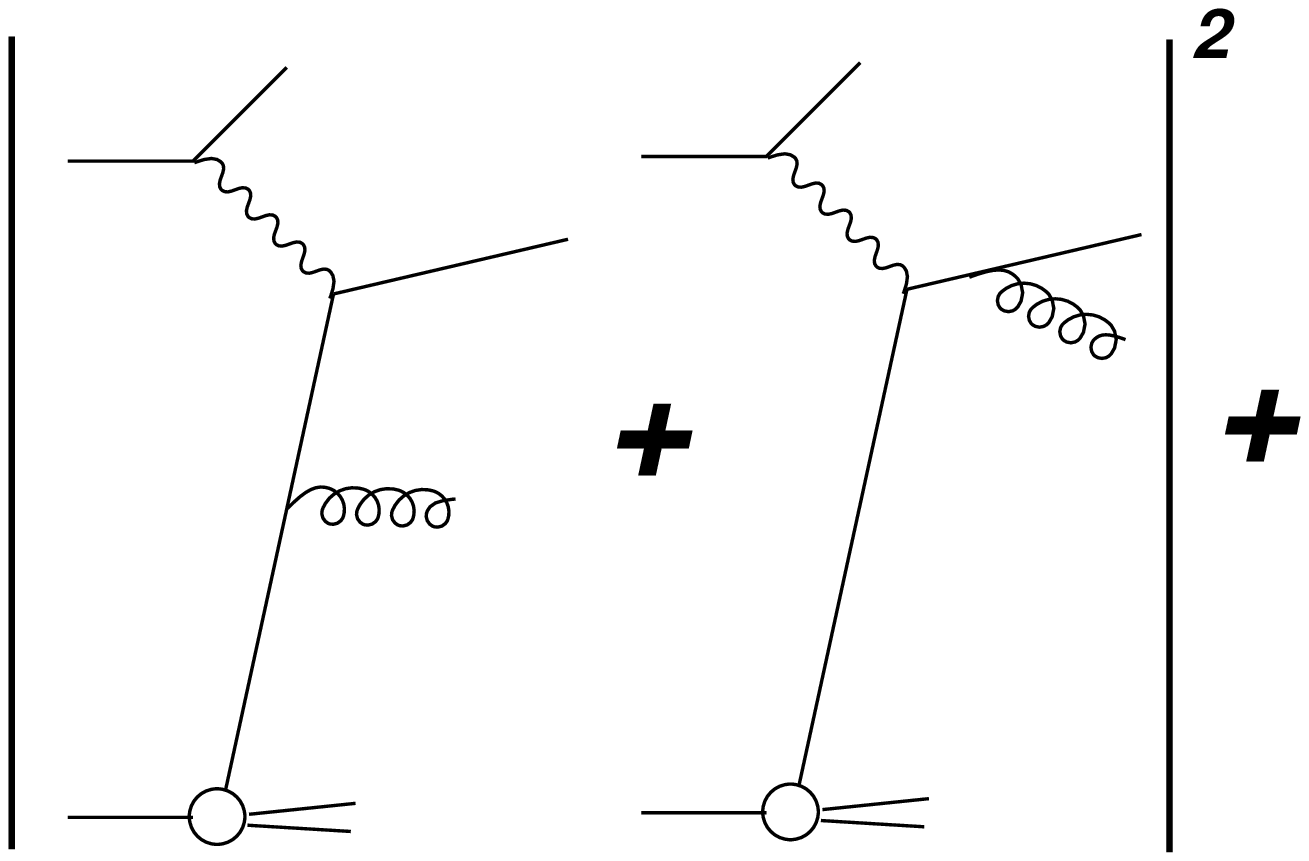}}}
\put(190,90){\mbox{\epsfxsize3.5cm\epsffile{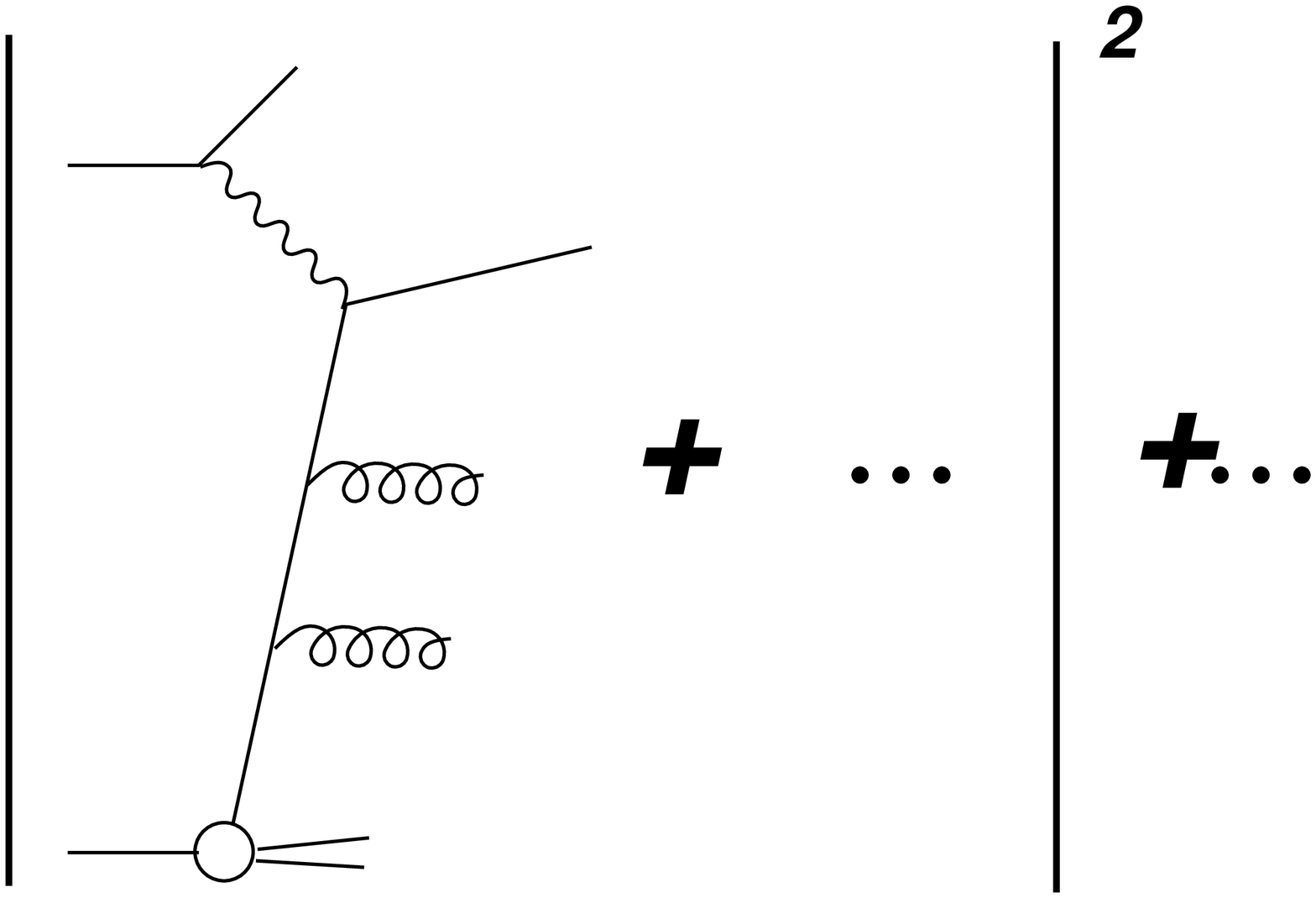}}}
\put(70,20){\mbox{\epsfxsize3.5cm\epsffile{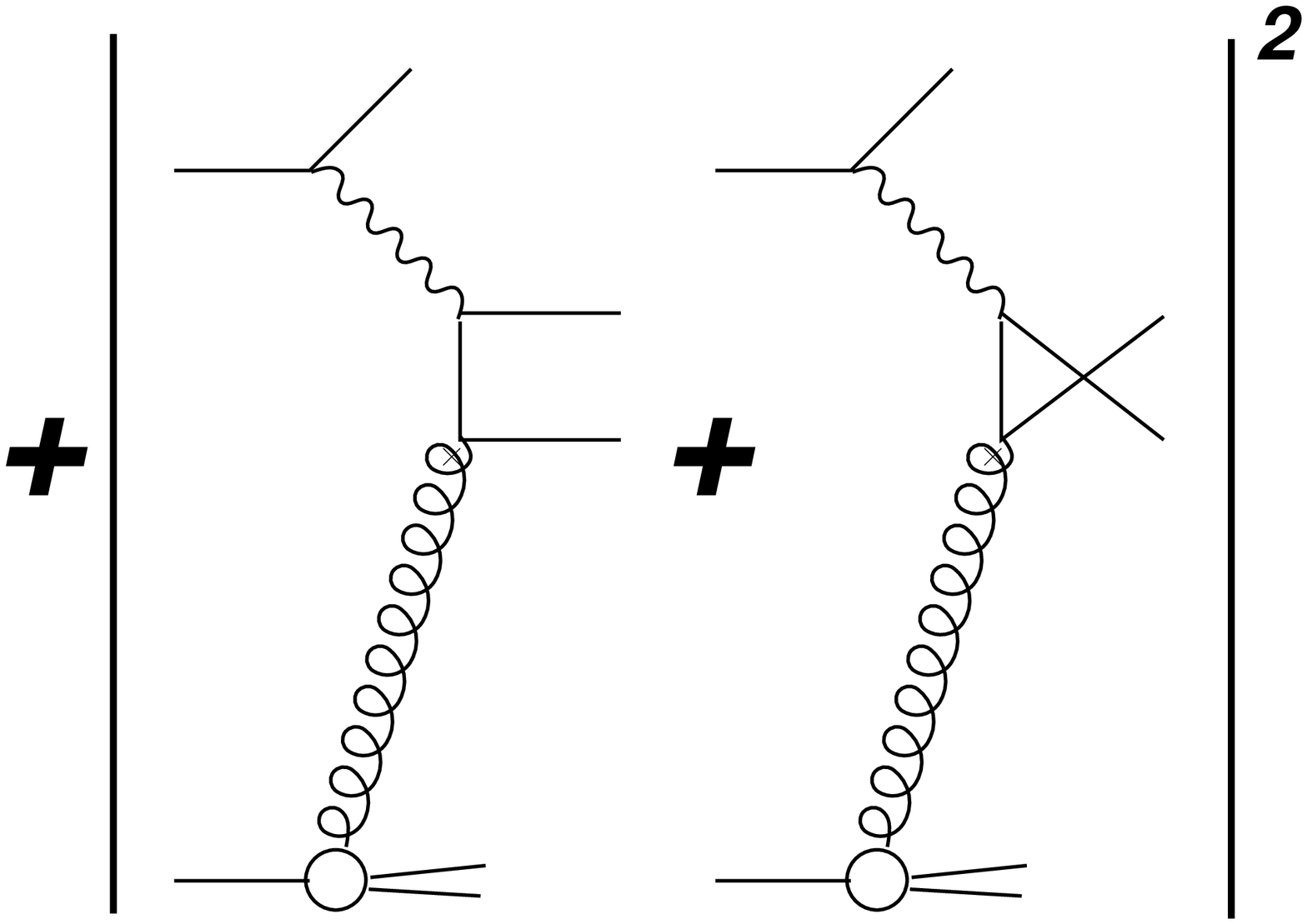}}}
\put(160,20){\mbox{\epsfxsize3.5cm\epsffile{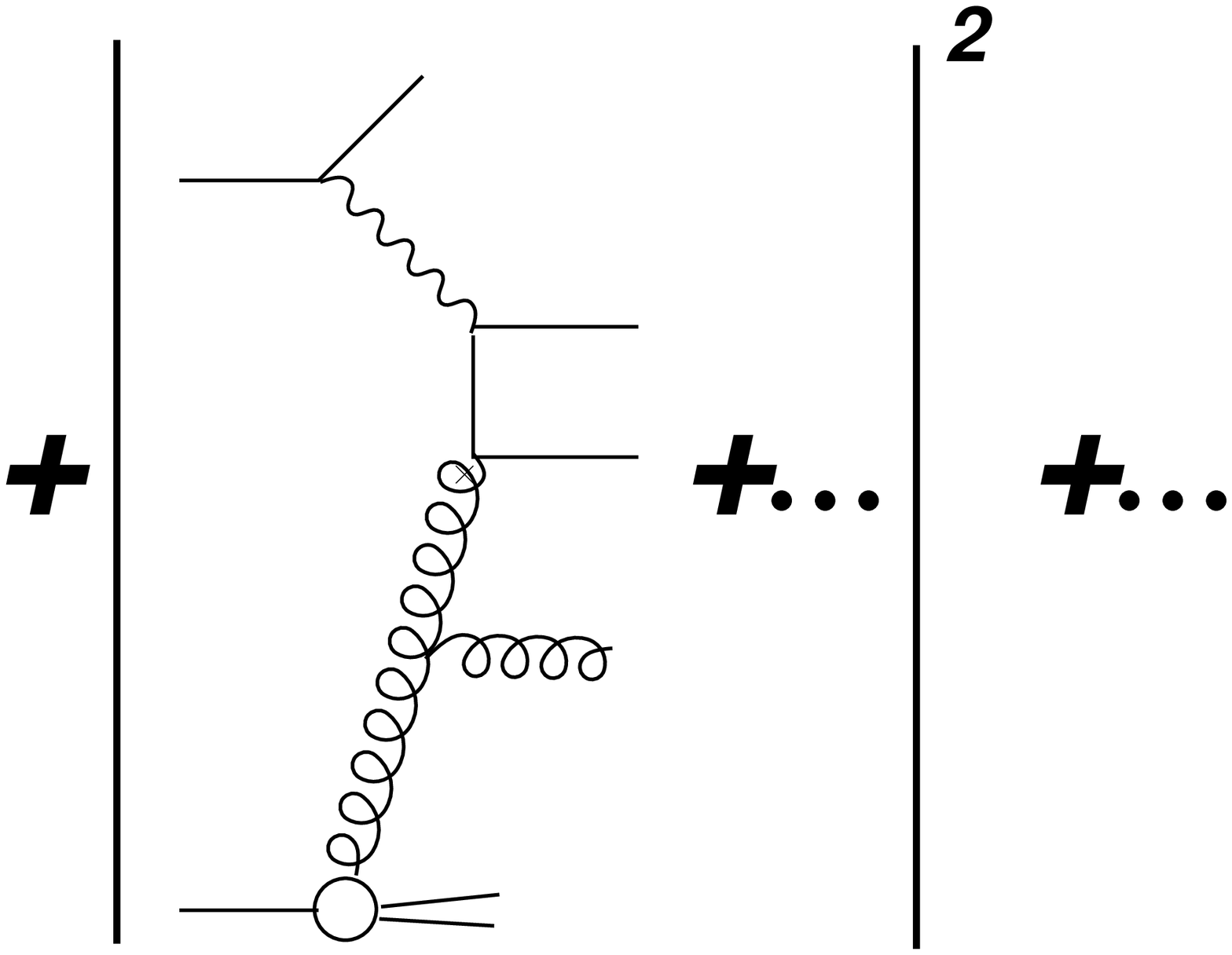}}}
\end{picture}
\caption{Lowest QCD corrections to $V^{\star}
\, q \rightarrow q$ \label{fig:OAS}}
\end{figure}

Because we are
considering the inclusive case, the calculation implies phase space integrations which lead to divergent integrals
when partons become collinear. They have to be regularized by
means of dimensional regularization \cite{Bol},
 a cut-off in the transverse momentum of the partons, or giving them masses.
 Any of these procedures introduce a regulator $\mu^2$ which has the dimension
 of the energy scale $Q^2$.

For example, to order $\alpha_s$, quarks contribute to the
structure function $F_2$ through the following convolution
integral
\begin{eqnarray}
F^{quarks}_2 && (x,Q^2)  =   \nonumber  \\ & & x\,
\sum_{f,\overline{f}} e_f^2\, \int_0^1\frac{dy}{y}\, q_f(y) \left[
\delta \left(1-\frac{x}{y} \right) + \frac{\alpha_s}{2\pi} \left\{
P_{qq}\left( \frac{x}{y} \right) \, \ln \frac{Q^2}{\mu^2} +
R\left(\frac{x}{y}\right) \right\} \right]
\end{eqnarray}
where $q_f(y)$ is the `bare' quark density. The delta function term comes
from the lowest order diagram in Figure \ref{fig:OAS}, and
represents the corresponding contribution to $F_2$ as in Equation
(25). The first term between the curly brackets includes the
factor  $\ln(Q^2/\mu^2)$ which diverges as the regulator $\mu^2$
goes to zero and is associated to the collinear gluon emission
diagrams, while the second term collects the finite contributions
from those diagrams. $P_{qq}(x/y)$ and $R(x/y)$ are known
calculable functions.

The collinear singularities of course threaten the validity of the
parton model, however they can be consistently removed from the
partonic subprocess absorbing them into the `bare' parton
densities $q_f(y)$ defining renormalized parton densities
\begin{eqnarray}
q_f(x,M^2)= q_f(x) +
\frac{\alpha_s}{2\pi}\int_0^1\frac{dy}{y}\,q_f(y) \left\{
P_{qq}\left( \frac{x}{y} \right) \, \ln \frac{M^2}{\mu^2} +
R'\left(\frac{x}{y}\right) \right\}
\end{eqnarray}
where $M^2$ is factorization scale chosen to separate the short
distance (`partonic') effects from the long distance (`hadronic')
ones. For the parton model to make sense, the renormalized parton
densities must be process independent, i.e. must be the same for
DIS, Drell-Yan, and any other process. Fortunately, this proves to
be the case to all orders of perturbation theory, and one ends up
with  an expression for the (physical, and thus finite) structure
function $F_2(x,Q^2)$ in terms of a finite renormalized parton
density $q_f(x,M^2)$ and an also finite partonic cross section
\begin{eqnarray}
F^{quarks}_2 && (x,Q^2)  =   \nonumber  \\ & & x\, \sum_{f,\overline{f}}
 e_f^2\, \int_0^1\frac{dy}{y}\,
q_f(y,M^2) \left[ \delta \left(1-\frac{x}{y} \right) + \alpha_s \,
C^q_2 \left(\frac{x}{y},\frac{Q^2}{M^2} \right)\right]
\end{eqnarray}
At this point, it is customary to chose the factorization scale
$M^2$ equal to the energy scale $Q^2$, factorizing the scale
dependence of the cross sections into the parton densities.

 The crucial observation here is that although perturbation
theory can not make an absolute prediction for $q_f(x,Q^2)$, from
Equation (25) it follows
\begin{eqnarray}
\frac{dq_i(x,M^2)}{d {\rm
log}Q^2}&=&\frac{\alpha_s}{2\pi}\int_{x}^{1}\frac{dy}{y} \left[
q_i(y,Q^2)\, P_{qq}\left(\frac{x}{y}\right) \right]
\end{eqnarray}
which means that QCD actually gives the $Q^2$ dependence of the
parton distributions.

By means of a similar procedure with the gluon densities, one can
deal with the divergences in the diagrams initiated by gluons.
However, due to the possibility of gluons emitting quark pairs
which then interact with the photon probe, the evolution the
evolution of the quark and gluon densities is no longer
independent, but coupled.
 Taking into account all the ${\cal{O}}(\alpha_s)$ contributions,
   we end up with the so called DGLAP equations
   (or Altarelli-Parisi equations for short)

\begin{eqnarray}
\frac{dq_i(x,Q^2)}{d \ln
Q^2}&=&\frac{\alpha_s}{2\pi}\int_{x}^{1}\frac{dy}{y} \left[
q_i(y,Q^2)\, P_{qq}\left(\frac{x}{y}\right) + g(y,Q^2)\,
P_{qg}\left(\frac{x}{y}\right) \right] \nonumber \\
\frac{dg(x,Q^2)}{d \ln
Q^2}&=&\frac{\alpha_s}{2\pi}\int_{x}^{1}\frac{dy}{y} \left[\sum_i
q_i(y,Q^2)\, P_{gq}\left(\frac{x}{y}\right) + g(y,Q^2)\,
P_{gg}\left(\frac{x}{y}\right) \right] \label{eq:ap}
\end{eqnarray}

Collecting all the ${\cal{O}}(\alpha_s)$ contributions to
$F_2(x,Q^2)$,  we  finally arrive to the most usual expression for
the structure function:
\begin{eqnarray}
F_2(x,Q^2) & = & x\, \sum_f e_f^2\, C_f(x)\otimes
\left[q_f(x,Q^2) + \bar{q}_f(x,Q^2) \right] \\ \nonumber &  & +
C_g(x) \otimes g(x,Q^2)
\end{eqnarray}
 where $\otimes$ indicates a
convolution integral and the index $g$ refers to gluons.

\begin{figure}[b]  
\centerline{\epsfig{file=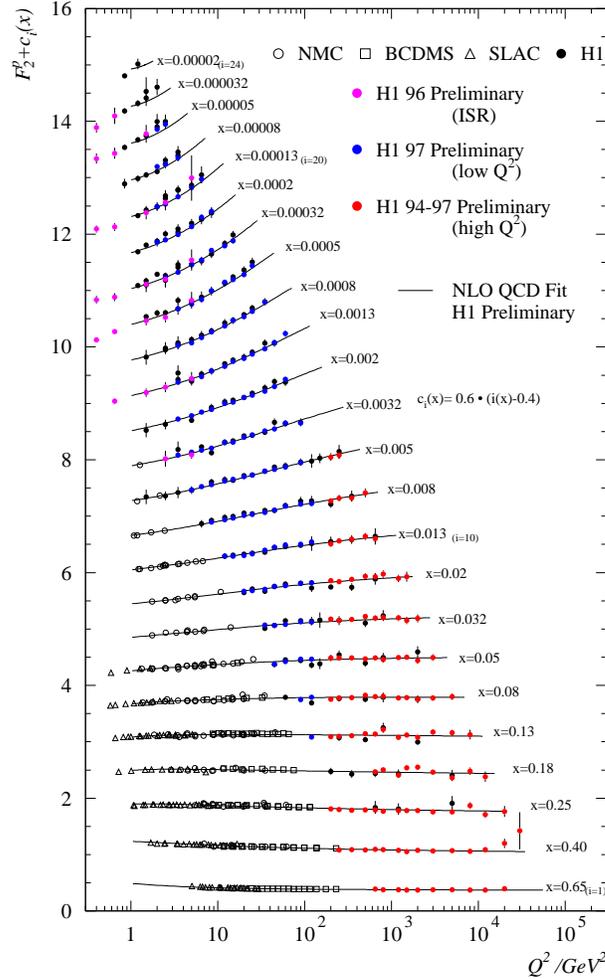,bbllx=115pt,bblly=55pt,
    bburx=485pt,bbury=780pt,width=6.5cm}}
\vspace{10pt} \caption{$F_2$ as a function of $Q^2$ and the
corresponding QCD fit \label{fig:F2HERA} }
\end{figure}

Deep Inelastic Scattering (DIS) experiments and hadron-hadron
interactions provide information on the parton structure of the
nucleon and constraints the dynamics of the quark-gluon
interaction given by QCD. In Figure \ref{fig:KDIS} the kinematic
regions in $x$ and $Q^2$ for cross section measurements in DIS
$ep$ scattering, $\nu$ scattering and jets in $p\bar{p}$
collisions were presented. Data from the last  years \cite{BASS}
have a clear influence on parton distributions and have motivated
the update of parton distribution function analysis. Thes most
recent sets are  CTEQ5 \cite{CTEQ5}, GRV98 \cite{GRV98} and
MRST \cite{MRST}. The data playing a fundamental paper in these
updates are the more precise ZEUS and H1 determinations of $F_2^p$
including $F_2^{charm}$; the NMC and CCFR final muon-nucleon and
neutrino data; the E866 $pp$ and $pd$ lepton pair production
asymmetry; the $W$ charge rapidity asymmetry; the D0 and CDF
analysis of inclusive single jet production and the E706 direct
photon production. It should be remarked that both approaches
differ in the selection of data for the global analysis that are
sensitive to gluons at large values of $x$. This means that a
better understanding of the gluon behaviour at large $x$ is in
order.

 The $F_2$ structure function measurements at HERA cover a
wide range of five orders of magnitude in both  $x$ and $Q^2$
values as shown in Figure \ref{fig:KDIS}. One important point
established by these data refers to the rise of the structure
function with decreasing $x$. Moreover, QCD fits including NLO
(next to leading order corrections) performed by the experimental
collaborations are in very good agreement with the data even at
low values of $Q^2 \simeq 1\,GeV$. In Figure \ref{fig:F2HERA},
$F_2$ is presented as a function of $Q^2$ for fixed values of $x$
from H1, ZEUS and the previous experiments, together with the QCD
fit \cite{BASS}.

\begin{figure}[t]  
\centerline{\epsfig{file=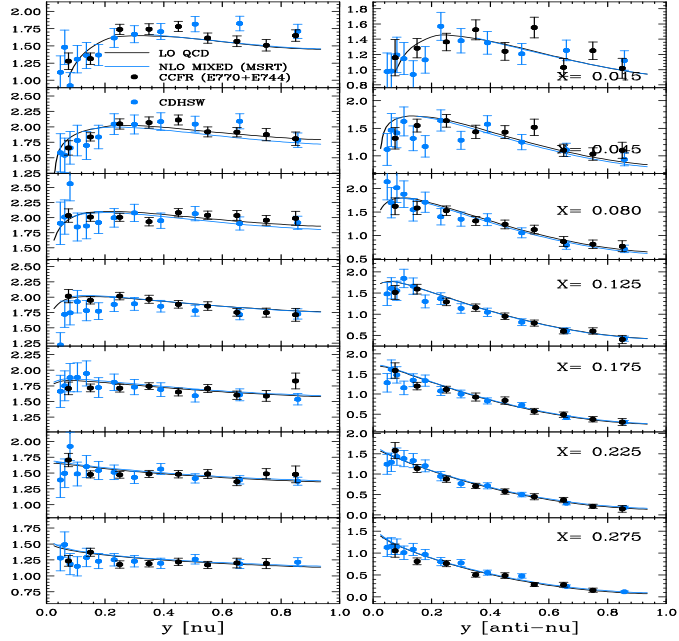,height=3.5in,width=3.5in}}
\vspace{10pt} \caption{$\nu$ and $\bar{\nu}$ DIS as a function of
$y$ from CCFR experiment together with QCD calculations
\label{fig:NUCCFR}}
\end{figure}

Figure \ref{fig:NUCCFR} presents de neutrino and antineutrino
cross section as a function of $y$ and different values of $x$ \cite{BASS}.
Data comes from the CCFR Fermilab experiment and is presented for
$\langle E_{\nu} \rangle = 150 \, GeV$. A good agreement with
previous data is observed and the NLO QCD analysis is
satisfactory.

We finish this paragraph with a brief comment about some QCD
technical details. The use of the renormalization group result for
$\alpha_s$ in the AP approach, specifically the replacement of
$\alpha_s$ by $\alpha_s(Q^2)$ in Equation (\ref{eq:ap}), has an
important consequence. For each order of perturbation in
$\alpha_s$ where the factorization procedure is performed, a whole
series of perturbative contributions is efectively re-summed, in
addition to the contributions comming form the corresponding
diagrams shown in Figure \ref{fig:OAS}. This kind of
contributions are depicted in Figure \ref{fig:LDD}, the so
called {\it ladder} diagrams.

\setlength{\unitlength}{1.mm} 
\begin{figure}[hbt]
\begin{picture}(150,75)(0,0)
\put(25,-85){\mbox{\epsfxsize12.cm\epsffile{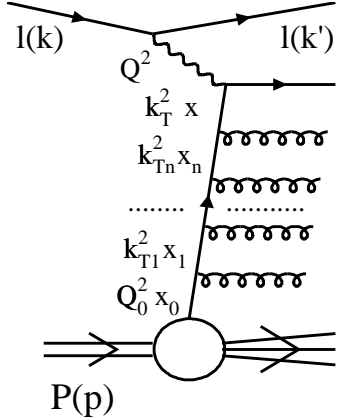}}} \caption{A
typical ladder diagram \label{fig:LDD}}
\end{picture}
\end{figure}

It can be shown that the improved AP approach takes into account
contributions coming from the sum of all ladder diagrams in which
the transverse momenta along the sides of the ladder are strongly
ordered, namely: $Q_0^2 \ll k_{T1}^2 \ll \cdots \ll k^2_{Ti}
\ll\cdots \ll Q^2$. This condition, of course implies also a
strong ordering for the momenta of the emitted partons.

Notice that the calculation of such a n-th order ladder diagram
implies integrations over the internal momenta. This nested
integration can be carried out because of the $k_T$ ordering
imposed, giving rise to  results proportional to powers of
$\alpha_s(Q^2)$ and $\ln \left(Q^2/Q_0^2\right)$.

Now, it is clear that large logarithms in $Q^2$ compensate the
small values of $\alpha_s(Q^2)$, that as we have seen, decreases
logarithmically. Consequently, all graph with rungs up to $n
\rightarrow \infty$ would have to be summed up. The so called {\it
leading log} approximation (LO), which is equivalent to follow the
renormalization group improved AP approach up to order $\alpha_s$,
collects all the contributions proportional to
\[
 \alpha_s^n(Q^2)\,\left[\ln \left(\frac{Q^2}{Q_0^2}\right)\right]^n
\]
while the {\it next to leading log} approximation (NLO),
equivalent to the second order extension of the AP approach, also
includes contributions proportional to
\[
 \alpha_s^n(Q^2)\,\left[\ln \left(\frac{Q^2}{Q_0^2}\right)\right]^{n-1}
\]

 The AP approximation is expected to be valid for $Q^2$ values sufficiently
 large but also for $x$ not too small, just to ensure that small values of $x$ do not give
rise to other large logarithms. In other words,
\[
\alpha_s(Q^2)\,\ln\left(\frac{1}{x}\right) \ll
\alpha_s(Q^2)\,\ln\left(\frac{Q^2}{Q^2_0}\right)< 1
\]

In is worth remarking that the AP equations can be solved
analytically when a strong ordering in $x$ is also requiered. In
this way one ends with the {\it double leading log approximation},
(DLL), where the large logarithmic terms are of the form \cite{Martin}
\[
 \alpha_s^n(Q^2)\,\left[\ln \left(\frac{Q^2}{Q_0^2}\right)\,
 \ln\left(\frac{1}{x}\right)\right]^n
\]
This approximation is expected to be valid for large $Q^2$ and
small $x$.

At small values of $x$, the parton content of the proton is gluon
dominated. In this case the AP DLL can be obtained with the result
\beq
 x\,g(x,Q^2) \approx x \,g_0(x,Q_0^2) \, exp
\sqrt{\frac{144}{25}\, \ln\left[\frac{\ln (Q^2/\Lambda^2)}{
\ln(Q_0^2/\Lambda^2)} \right]\,\ln(1/x)}  \label{eq:DLL}
 \eeq
that shows that the gluon density increases faster than a power of
$\ln (1/x)$.

 In some cases of interest, $x$ is small enough
but $Q^2$ is not sufficiently large to be inside the DLL regime.
Under these circumstances, the AP approximation is not more valid.
The BFKL (Balitsky, Fadin, Kuraev, Lipatov) equation has been
proposed to tackle the limit behaviour of large $1/x$ and $Q^2$
finite and fixed. In this scheme, the $x_i$ variables in the
ladder are stongly ordered, namely: $x_0 \ll x_1 \ll \cdots \ll
x_i \ll \cdots \ll x$; while there is no order on $k_T$ imposed.
This approach ends with the so called leading log approximation in
$\ln(1/x)$. The region of validity being
\[
\alpha_s(Q^2)\,\ln\left(\frac{Q^2}{Q_0^2}\right) \ll
\alpha_s(Q^2)\, \ln \left(\frac{1}{x}\right) < 1
\]

Due to technical reasons, the BFKL equation is written in terms of
the function $f(x,k_T^2)$, related to the usual gluon density
$g(x,Q^2)$ that dominates at very small $x$ by \beq
 x\,g(x,Q^2) =
\int^{Q^2}_0 dk_T^2\,\frac{f(x,k_T^2)}{k_T^2}
 \eeq
The standard form of the equation is \beq \frac{\partial
f(x,k_T^2)}{\partial \ln(1/x)} =
\frac{3\,\alpha_s}{\pi}\,k_T^2\,\int_0^{\infty}
\frac{dk^{\prime2}_T}{k_T^{\prime 2}}\,\left[\frac{f(x,k_T^{\prime
2})- f(x,k_T^2)}{|k^{\prime 2}_T - k^2_T|} +
\frac{f(x,k_T^2)}{\sqrt{4\,k^{\prime 4}_T + k_T^4}} \right]
 \eeq

This approximate BFKL equation can be solved analytically for
fixed $\alpha_s$. The solution behaves like
\[
f(x,k_T^2) \propto \left(\frac{x}{x_0}\right)^{- \lambda}
\]
with \beq
 \lambda = \frac{N_c\,\alpha_s}{\pi}\,4\,\ln 2 \approx
0.5
 \eeq
  for $N_c = 3$ and $\alpha_s = 0.19$.

Consequently, the gluon density rises like a power of $1/x$ for
decreasing $x$, namely
\[
x\,g(x,Q^2) \propto x^{- \lambda}
\]
clearly faster than the AP DLL prediction (\ref{eq:DLL}). It
should be noticed that a running $\alpha_s$ and higher order
corrections decrease the value of $\lambda$.

\setlength{\unitlength}{1.mm} 
\begin{figure}[t]
\begin{picture}(150,80)(0,0)
\put(80,11){\mbox{\epsfxsize5.cm\epsffile{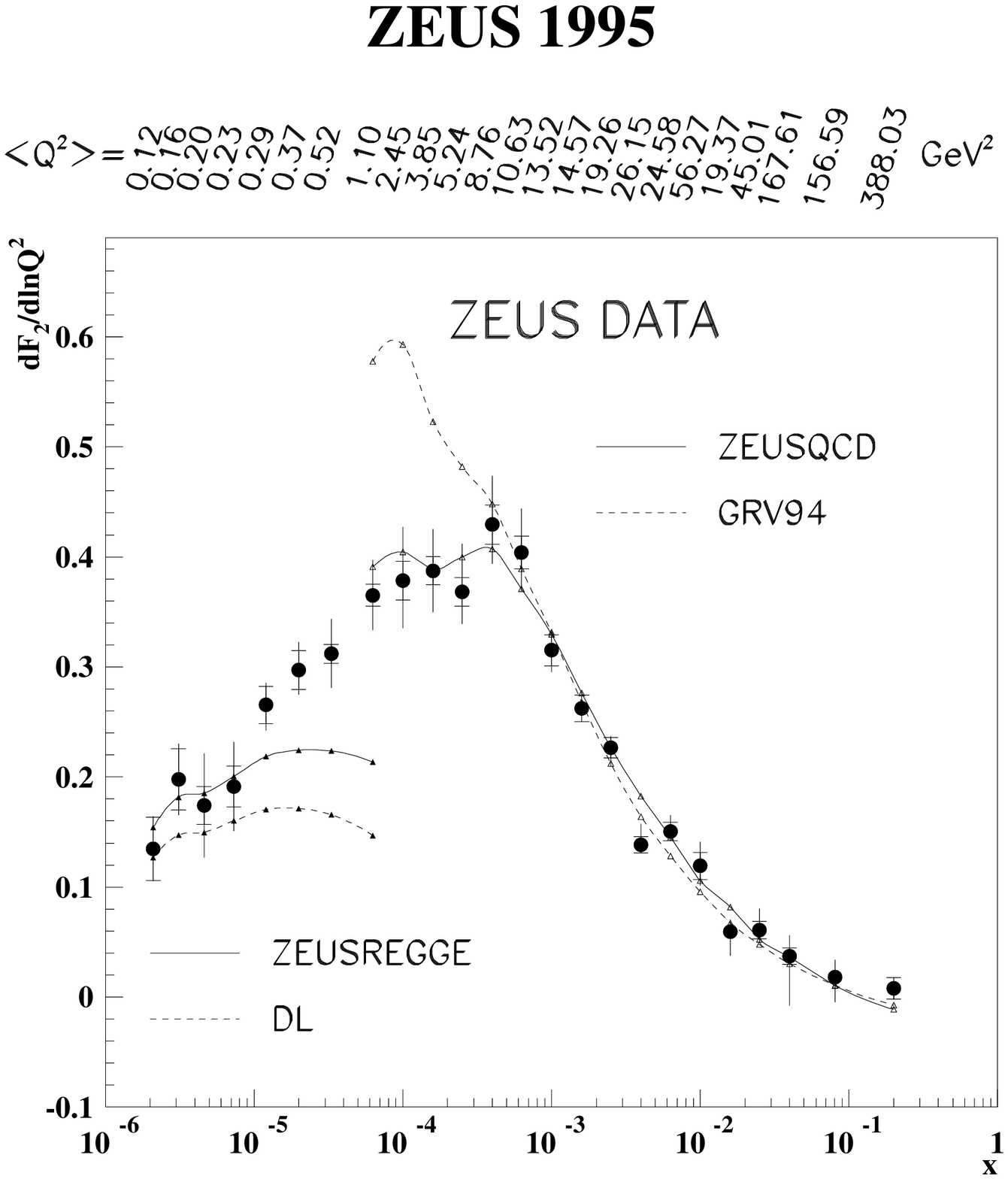}}}
\put(15,10){\mbox{\epsfxsize5.cm\epsffile{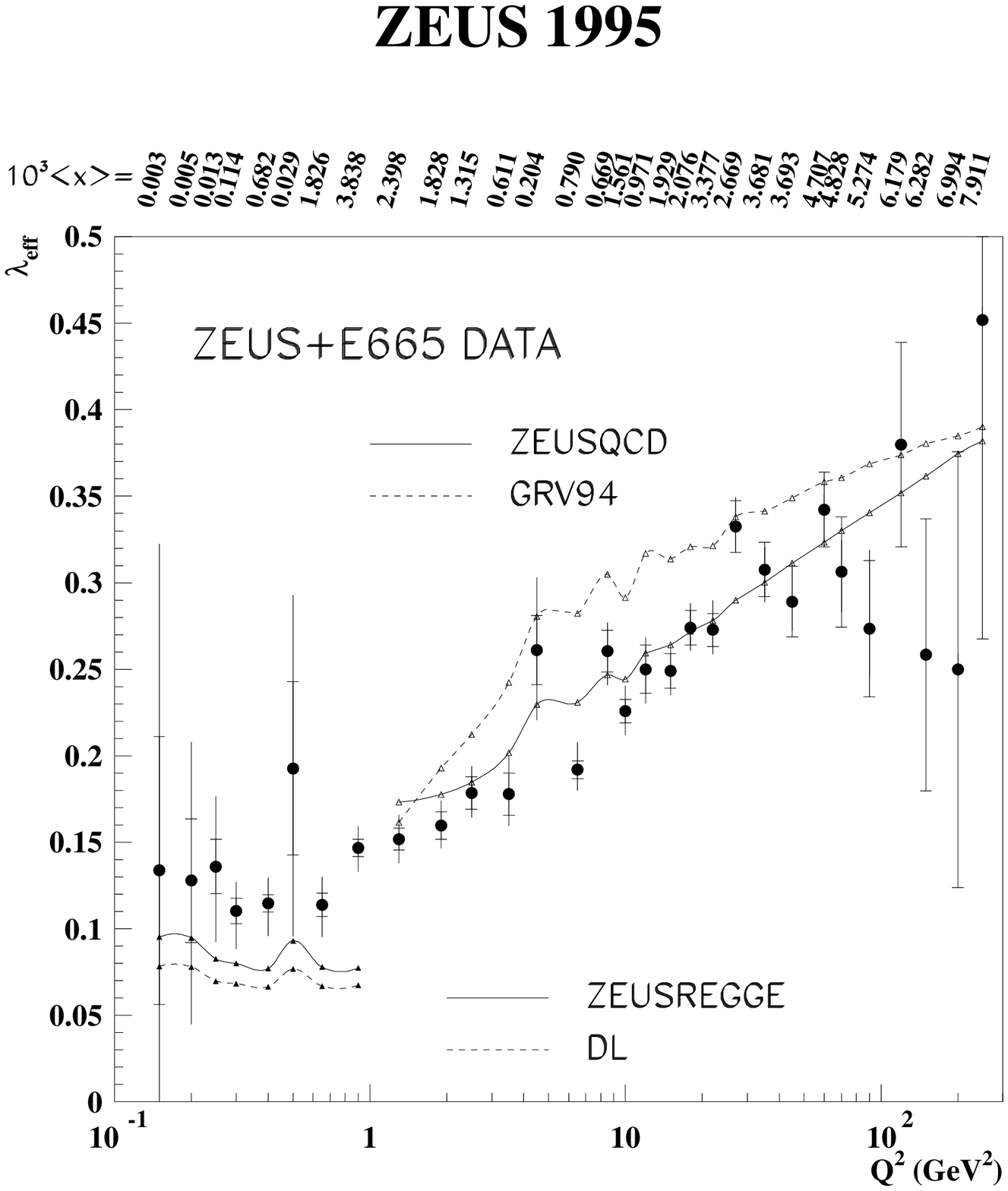}}}
\caption{$\lambda_{eff}$ and $dF_2/d\mbox{ln} Q^2$.
 \label{fig:xlambda} }
\end{picture}
\end{figure}

Assuming that gluons dominate at low
$x$ one can expect an approximate behaviour for the structure function given by
\[  F_2(x,Q^2) \mid_{Q^2} =  c\, x^{- \lambda_{eff}} \mid_{Q^2} \]
Figure \ref{fig:xlambda} (a) shows the values obtained for
$\lambda_{eff}$ as a function of $Q^2$ from fits to ZEUS and E665
data. Figure \ref{fig:xlambda} (b) shows $dF_2/d\mbox{ln} Q^2$ as
a function of $x$ \cite{9812029}. In both figures QCD and Regge
inspired fits have been included. The schematic map of the plane
$\ln(Q^2), \ln(1/x)$ shown in Figure \ref{fig:Q2x} intents to
divide the regions where each of the two approaches, DGLAP and
BFKL, are in order \cite{9712505}.

\setlength{\unitlength}{1.mm} 
\begin{figure}[hbt]
\begin{picture}(150,70)(0,0)
\put(35,12){\mbox{\epsfxsize8.cm\epsffile{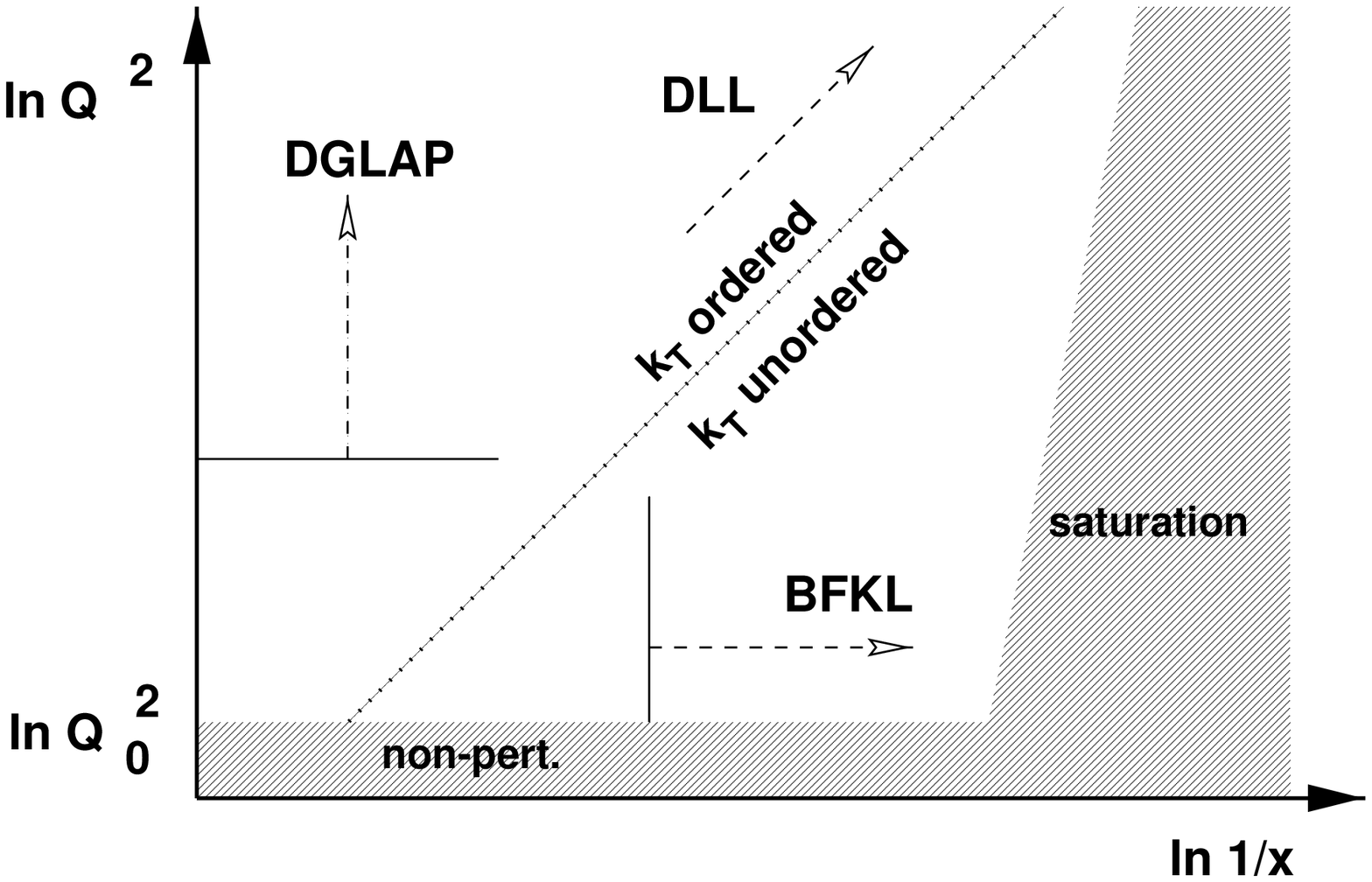}}} \caption{The
schematic map of the plane $\ln(Q^2), \ln(1/x)$ \label{fig:Q2x} }
\end{picture}
\end{figure}

In the figure, a {\it saturation} region is indicated, where one
waits the gluon density to be so high as to prevent its continuous
growing.

\subsection*{Semi-inclusive}

In the analysis of hadron structure, more information can be
obtained if one goes one step beyond totally inclusive DIS, namely
to semi-inclusive DIS, where one of the final hadrons is also
measured
\[
\ell + h \rightarrow \ell^{\prime} + h^{\prime} + X
\]

In describing semi-inclusive processes, in addition to quark
distributions, the so called {\it fragmentation functions} are
necesary. These functions \cite{FEY} describe, or better
parametrize, a given parton decay into a final hadron.

It is clear that, in principle, the best process to analyze
fragmentation functions is the semi-inclusive $e^+\,e^-$
annihilation, when a single hadron is fully detected in the final
state
\[
e^+\,e^- \rightarrow h(P) + X
\]
The corresponding kinematics is depicted in Figure \ref{fig:FRA}.

\setlength{\unitlength}{1.mm} 
\begin{figure}[hbt]
\begin{picture}(150,60)(0,0)
\put(15,-92){\mbox{\epsfxsize12.cm\epsffile{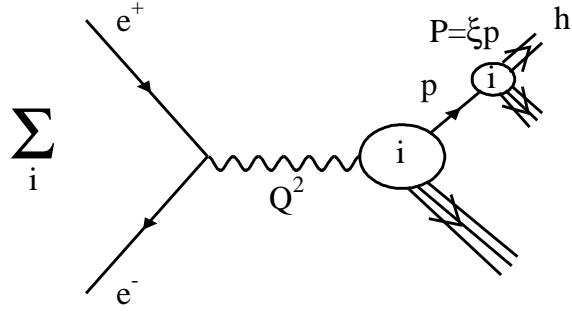}}}
\vspace{10pt}
 \caption{Schematic view  of semi-inclusive $e^+\,e_-$
\label{fig:FRA}}
\end{picture}
\end{figure}

The relevant variables for this case are \beq z =
\frac{2\,P\,q}{Q^2} \,\,\,\,\mbox{and}\,\,\,\, \tilde{z}= \frac{2\,p\,q}{Q^2}
\eeq

It is also convenient to introduce $\zeta$ that measures the
fraction of the parton momentum carried by the final hadron. The
corresponding cross section reads
 \begin{eqnarray}
\frac{d\,\sigma_h (z,Q^2)}{dz\,dQ^2} & = &
\frac{4\,\pi\,\alpha^2}{Q^2} \sum_i \int  d\tilde{z}\,\int
d\zeta\,\delta(z -
\tilde{z}\,\zeta)\,\tilde{\sigma}^i_p(\tilde{z},Q^2)\,\zeta\,D^h_i(\zeta,Q^2)\\
\nonumber & = & \sum_i \int_z^1 \frac{d\zeta}{\zeta}\,
\tilde{\sigma}^i_p \left(\frac{\tilde{z}}{\zeta},Q^2 \right)\,
\zeta\,D^h_i(\zeta,Q^2)
\end{eqnarray}

Here $\tilde{\sigma}^i_p$ is the parton $i$ production
cross-section, that in the parton model results

\beq \tilde{\sigma}^i_p(\tilde{z},Q^2)= e_i^2\,\delta(1-\tilde{z})
\eeq if $i=q,\,\bar{q}$ and vanishes if $i=g$. The function
$D^h_i(\zeta,Q^2)$ is the {\it fragmentation function} or
probability for a parton $i$ to decay into a hadron $h$ carrying
the fraction $\zeta$ of the parton momentum.

When QCD corrections are in order, the fragmentation functions
acquire a $Q^2$ dependence. The evolution of these functions with
the scale \beq t= \ln \left(\frac{Q^2}{\mu^2} \right) \eeq is
given by the AP-like  integrodifferential equations
\begin{eqnarray}
\frac{dD^h_{q}}{dt}(z,t) & = &
\frac{\alpha_s(Q^2)}{2\,\pi}\,\int_z^1
\frac{dy}{y}\,\left[D^h_{q}{dt}(y,t)\,P_{qq}(z/y) + 2\,f\,
D^h_{g}{dt}(y,t)\,P_{gq}(z/y)\right] \\
 \frac{dD^h_{g}}{dt}(z,t) &
= & \frac{\alpha_s(Q^2)}{2\,\pi}\,\int_z^1
\frac{dy}{y}\,\left[D^h_{q}{dt}(y,t)\,P_{qg}(z/y) +
D^h_{g}{dt}(y,t)\,P_{gg}(z/y)\right]
\end{eqnarray}

The functions $P_{ij}$ are exactly the same as those appearing in
the AP equations for the structure functions (\ref{eq:ap}). The
corresponding graphical interpretation in terms of the basic QCD
processes is included in Figure \ref{fig:APF}.

\setlength{\unitlength}{1.mm} 
\begin{figure}[hbt]
\begin{picture}(150,50)(0,0)
\put(64,-95){\mbox{\epsfxsize12.5cm\epsffile{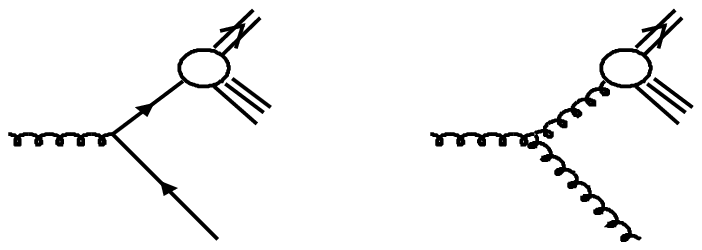}}}
\put(-2,-95){\mbox{\epsfxsize12.5cm\epsffile{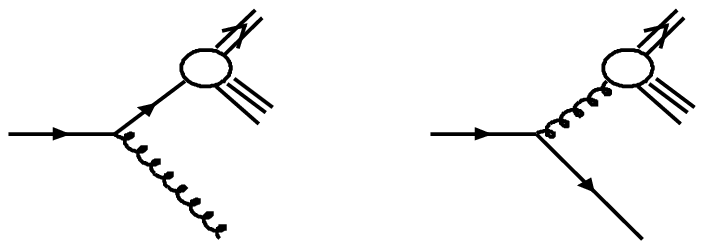}}}
\vspace{10pt}
\caption{QCD processes providing the evolution of fragmentation functions}
\label{fig:APF}
\end{picture}
\end{figure}

It is worth noticing that the properties of the splitting
functions are enough to guarantee the momentum sum rule \beq
\sum_h\,\int_0^1 z\,D^h_{q}(z,Q^2) = 1 \eeq and the analogous one
for the gluon fragmentation  function.

To solve the AP equations it is necessary to know the
fragmentation functions at some value $Q^2_0$. Exactly as in the
case of structure functions, they have to be obtained from
experimental data. In general one proposes a given form for the
initial $D$-function, in agreement with data, and then AP
equations provide the evolution with $Q^2$.

\subsection*{Fracture Functions}

For those who are not acquainted with fracture functions, let us
briefly summarize that the main idea behind fracture functions is
the realization that the most familiar perturbative description
for semi-inclusive processes, based on parton distributions and
fragmentation functions is, at least, incomplete \cite{VEN}.

In the usual approach to  semi-inclusive DIS, the corresponding
cross section is
expressed by a convolution between parton distributions and
fragmentation functions accounting the process in which  the
struck parton hadronizes into a detected final state particle.
\begin{equation}
\frac{d^3\sigma_{curr.}}{dx\,dy\,dz} \simeq \frac{4\pi
\alpha^2}{x\,(p+k)^2}\frac{1+(1-y)^2}{2y^2} \sum_i
e^2_i\,q_i(x,Q^2) \times D^h_i (z,Q^2)
\end{equation}
Obviously, this approach only takes into account hadrons produced
in the current fragmentation region, and  what is more, within this
approximation, and  in leading order, hadrons can only be produced
in the backward direction. Going to higher orders one finds a
breakdown of hard factorization, as there are collinear
singularities  that can not be substracted in the parton
distributions or in the fragmentation functions.

All this means
that there are additional contributions missing, which are mainly
target fragmentation processes and are included in the so called fracture functions $M_i^h$, defined by
\begin{equation}
\frac{d^3\sigma_{targ.}}{dx\,dy\,dz} \simeq  \frac{4\pi
\alpha^2}{x\,(p+k)^2}\frac{1+(1-y)^2}{2y^2} \sum_i e^2_i\,  (1-x)
\,M^h_i(x,z,Q^2)
\end{equation}
These functions, that can be thought as the probabilities to find
a parton of a given flavour in an already fragmentated target,
straightforwardly solve the factorization problem and also allow a
LO description of hadrons produced in the forward direction. The
only subtlety regarding them is that they obey slightly different
evolution equations; their scale dependence not only depends on
their shape at a given scale but also on that of ordinary
structure functions and fragmentation functions, reflecting the
fact that current and target fragmentation are not truly
independent of each other. This is usually refered to as `non
homogeneus' evolution
\begin{eqnarray}
\frac{\partial M^h_i(x,z,Q^2)}{\partial \mbox{ln}Q^2}
=\frac{\alpha_s(Q^2)}{2\pi}\int \frac{dy}{y} P_{ij}(y)
M^h_j(\frac{x}{y},z,Q^2)\\ +  \frac{\alpha_s(Q^2)}{2\pi}\int
\frac{dy}{x(1-y)} \hat{P}_{ijl}(y)q_j(\frac{x}{y},Q^2)
D^h_l(\frac{zy}{x(1-y)},Q^2)  \nonumber
\end{eqnarray}
As for structure functions in totally inclusive deep inelastic
scattering, QCD does not predict the shape of fracture functions
unless it is known at a given initial scale. This nonperturbative
information have to be obtained from the experiment, and,
eventually, can be parametrized finding inspiration in
nonperturbative models, as it is the case for ordinary structure and fragmentation functions.

Fracture functions have been succesfully applied to the description of
leading baryon production \cite{LB}, have been extended to spin dependent
processes \cite{FRACPOL}, and as we shall see, are relevant for the
description of diffractive DIS.

\section*{Photon}

It may seem strange that the next step in our study of parton
structure relies in a particle, the photon,  that it is not a
hadron and of course has no structure by itself, as we very well
know. However, a high energy photon can `develop' structure when
interacting with another object. In fact, as the electromagnetic
field couples to all particles carrying the electromagnetic
current, a photon can fluctuate into particle-antiparticle virtual
states, particularly into quark-antiquark pairs or even more complex
hadronic objects with the same quantum numbers. As long a the
fluctuation time is longer than the interaction time we can talk
about the structure of the the photon, and deal with it as a `very
special' hadron. Notice that at high energies, the fluctuation of
a photon into a state of invariant mass $M$ can persist for a time
of the order of $\tau \sim 2\,E_{\gamma}/M$ until the virtual
state materializes by a collision or annihilation with another
system \cite{BRZ}.

\setlength{\unitlength}{1.mm} 
\begin{figure}[hbt]
\begin{picture}(150,60)(0,0)
\put(33,-66){\mbox{\epsfxsize10.cm\epsffile{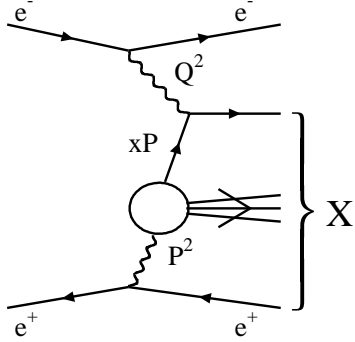}}}
\caption{Photon Structure.}
\label{fig:e+e-}
\end{picture}
\end{figure}

There are many reasons for this peculiar hadronic structure to be
considered `matter' of study. First of all, it gives the
opportunity to study how QCD works in a completely different
scenario. Theoretical studies of photonic parton distributions of
real, i.e. on shell, photons have a long history initiated by
Witten's work \cite{witten}. This description have been well
established through measurements in $\gamma \gamma$ collisions at
$e^+ e^-$ colliders (PETRA, PEP, LEP) however, extending the idea
of photon structure to high virtualities as in $ep$ processes like
dijet production, a new insight has been gained as we link DIS,
photoproduction, and $\gamma \gamma$ interactions \cite{Fer}.

The concept of photon structure functions for real and virtual
photons can be defined and understood in close analogy to deep
inelastic lepton nucleon scattering, via the subproccess
$\gamma^*(Q^2)\gamma(P^2) \longrightarrow X$  in $e^+e^-
\longrightarrow  e^{\pm}X$ as in Figure \ref{fig:e+e-},
where we denote the probed target photon with virtuality
$P^2=-p^2_{\gamma}$ by $\gamma(P^2)$ and reserve $\gamma^*(Q^2)$
for the highly virtual one. The relevant
differential cross section can be expressed, as in the hadronic case, in
terms of the usual scaling variables $x$ and $y$ as
\begin{eqnarray} \label{eq:photon}
\frac{d^2\sigma (e\gamma(P^2)\longrightarrow eX)}{dxdy} & = &
\frac{2\,\pi\, \alpha^2\, s_{e\gamma}}{Q^4}\left[
(1+(1-y)^2)F_2^{\gamma(P^2)}(x,Q^2) \right.
 \\ \nonumber & &
\left. -y^2F_L^{\gamma(P^2)} (x,Q^2) \right]
\end{eqnarray}
with $F_{2,L}^{\gamma(P^2)}(x,Q^2)$ denoting the photonic
structure functions. The measured $e^+e^-$ cross section is
obtained by convoluting Equation(\ref{eq:photon}) with the photon
flux for the target photon $\gamma(P^2)$ \cite{wei}. The range of
photon virtualities explored in the above mentioned experiment is
given by
\begin{equation}
m_e^2\,y^2/(1-y) \leq P^2_{min} \leq P^2 \leq P^2_{max} \leq
\frac{s_{e\gamma}}{2} (1-y) (1- \cos \theta_{max})
\end{equation}
where $m_e$ is the mass of the electron, $s_{e\gamma}$ is the
square of the c.m.s. energy, $y$ is the energy fraction taken by
the photon, and $\theta_{max}$ is the maximum
 scattering angle of the electron in the c.m.s. frame. $P^2_{min,max}$ are further
  determined by detector specifications and experimental settings.

\setlength{\unitlength}{1.mm} 
\begin{figure}[hbt]
\begin{picture}(150,55)(0,0)
\put(7,-60){\mbox{\epsfxsize9.cm\epsffile{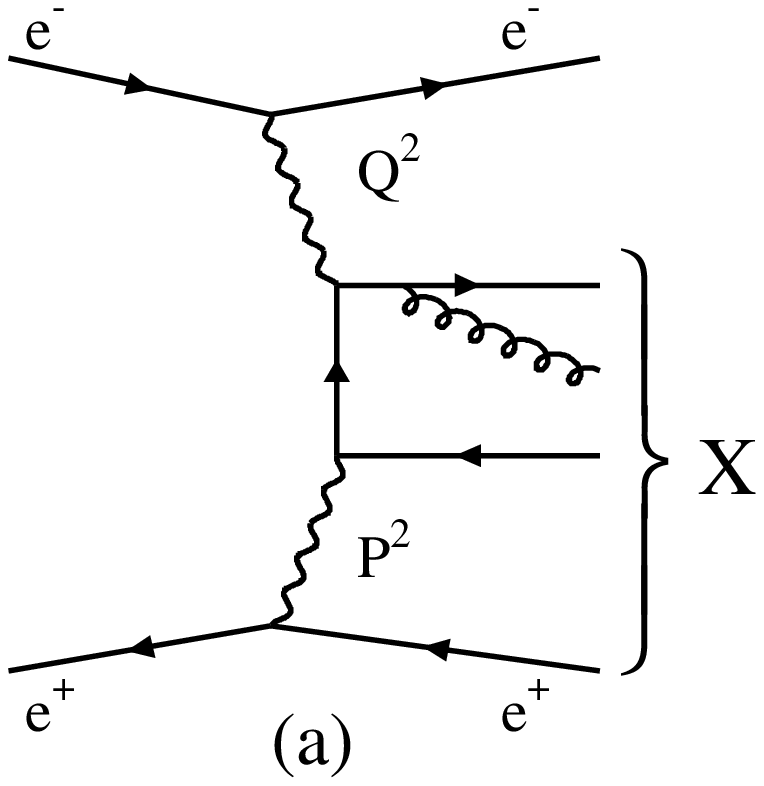}}}
\put(67,-60){\mbox{\epsfxsize9.cm\epsffile{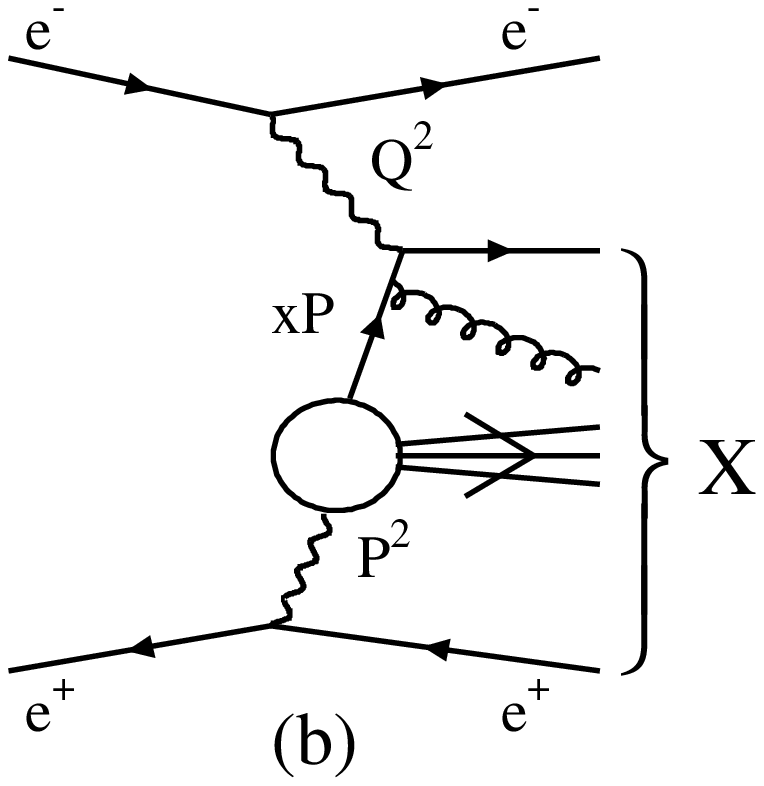}}}
\caption{Direct (a) and Resolved (b) Contributions to $F_{2,L}^{\gamma(P^2)}(x,Q^2)$.}
\label{fig:dires}
\end{picture}
\end{figure}

It is worthwhile noticing that if one could neglect photon
fluctuations into complex hadronic states, then the dependence of
$F_{2,L}^{\gamma(P^2)}(x,Q^2)$ in both $x$ and $Q^2$ would be
fully predictable in perturbative QCD. The pointlike process
$\gamma^*(Q^2)\gamma(P^2)\longrightarrow q\bar{q}$, corrected by
gluon radiation effects, yields a definite prediction for
$F_{2,L}^{\gamma(P^2)}(x,Q^2)$. However, as it was anticipated,
this is only the so called `direct' contribution to the total
process, named in opposition to the `resolved' one, where the
virtual photon $\gamma^*(Q^2)$ strikes the nonperturbative
`dressing' of the photon fluctuation (Fig.\ref{fig:dires}).

\setlength{\unitlength}{1.mm} 
\begin{figure}[b]
\begin{picture}(150,88)(0,0)
\put(30,10){\mbox{\epsfxsize9.5cm\epsffile{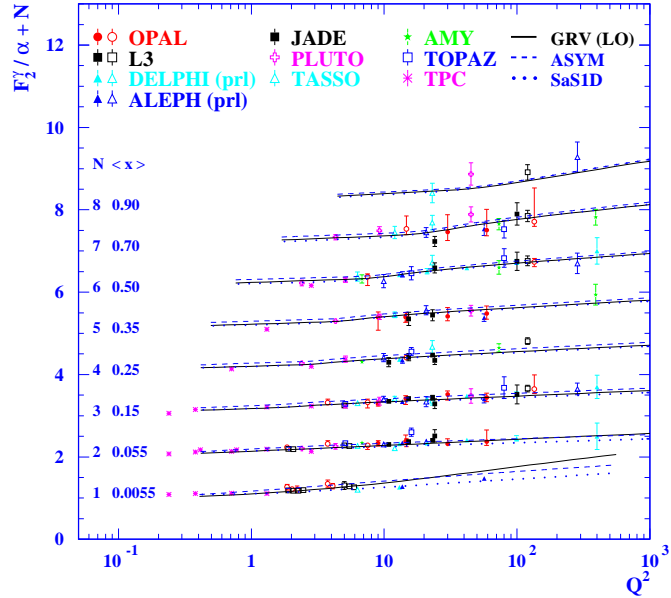}}}
\caption{ $F_{2}^{\gamma}(x,Q^2)$ as a function of $Q^2$.}
\label{fig:phQ^2}
\end{picture}
\end{figure}

QCD corrections to the `direct' component not only take into
account the splitting of partons into partons as in the ordinary
AP evolution equations, but also the possibility of a photon to
split into a quark-antiquark pair. This implies an inhomogeneous
term in the evolution equations and leads to a logarithmic
enhancement in $F_{2,L}^{\gamma(P^2)}(x,Q^2)$ which means positive
scaling violations for all values of $x$ (Fig.\ref{fig:phQ^2}) \cite{9812030b}.

The `resolved' component obeys the same evolution equations typical of the
 lepton-hadron interactions. In a Vector Meson Dominance approach for the
  photon fluctuations, the resolved component is expected to vanish like
   $(1/P^2)^2$ in connection
to the vector meson propagator, however it is not clear up to which values
 of $P^2$ the nonperturbative contributions are relevant.
In recent years several sets of parton distributions for real and
virtual photons have been proposed \cite{glu,phpd}. For virtual
photons, different
 approaches have been followed in LO y NLO global fits to the available LEP data
(Figure \ref{fig:f2g}) \cite{9812030b}.

\setlength{\unitlength}{1.mm} 
\begin{figure}[hbt]
\begin{picture}(150,110)(0,0)
\put(25,-10){\mbox{\epsfxsize9.5cm\epsffile{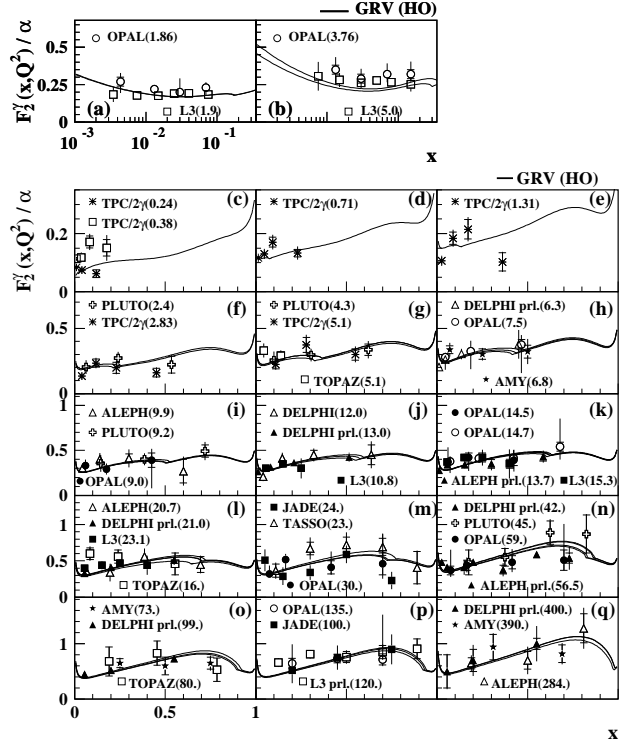}}}
\caption{ $F_{2}^{\gamma}(x,Q^2)$ as a function of $x$.}
\label{fig:f2g}
\end{picture}
\end{figure}

LEP data constraint reasonably well the quark distributions in the photon,
 however, in order to constrain the gluon densitiy it is much more suitable
 to analyze dijet production processes in $e^+ \,p$ collisions as measured at HERA.
As we shall see, in this context the picture of `direct' and `resolved'  is
 also more clearly illustrated. In this kind of process the parton content of
 the proton is used to probe the iqi partonic structure of the photon as in a
  hadron-hadron collision. The hard scale invoked to allow such inspection
   and guarantee the validity of
 a perturbative treatment, is the large transverse momentum of the final state jets.
   Figure \ref{fig:dijet} depicts LO `direct' and `resolved' contributions to
   dijet production $e^+ \,p$ collisions.
As it can be seen, even at the lowest order, photonic gluons
contribute to the cross section, at variance with what happens in
$e^+ e^-$ collisions.

\setlength{\unitlength}{1.mm} 
\begin{figure}[hbt]
\begin{picture}(150,55)(0,0)
\put(7,-60){\mbox{\epsfxsize9.cm\epsffile{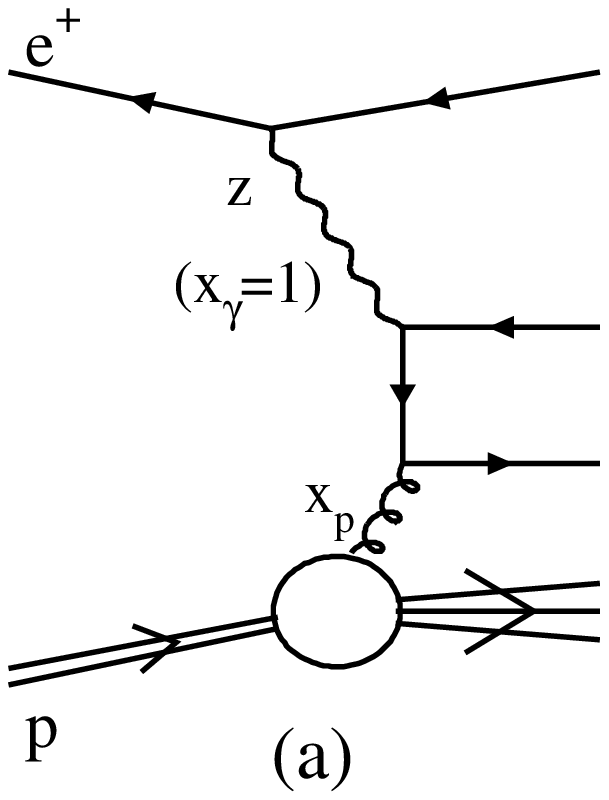}}}
\put(67,-60){\mbox{\epsfxsize9.cm\epsffile{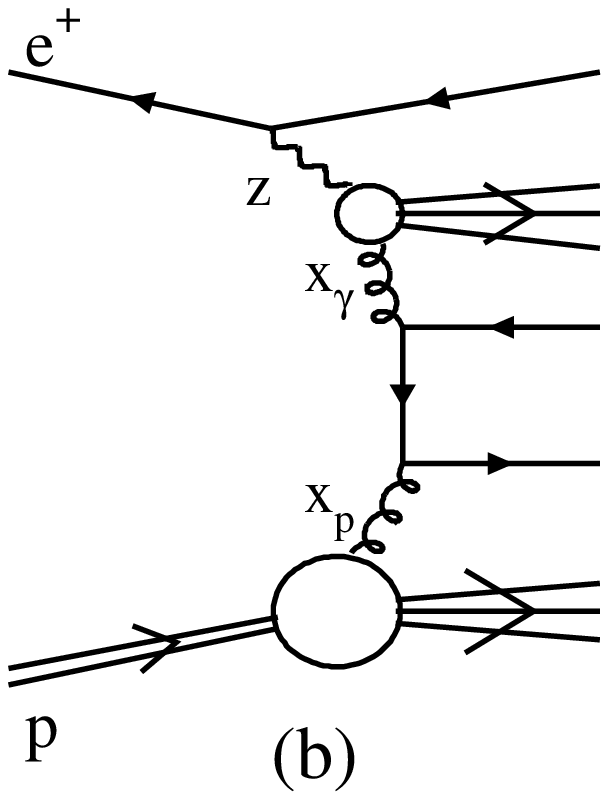}}}
\caption{Direct (a) and Resolved (b) Contributions to dijet production.}
\label{fig:dijet}
\end{picture}
\end{figure}

In leading order the differential cross section for two jet production in $ep$
 collisions takes
a very simple form  when written in terms of the fraction of the photon energy
 intervening in the
hard process, $x_{\gamma}$, the fraction of the proton energy carried by the participating
parton, $x_p$, and that of the electron carried by the photon, $z$ \cite{DG0}
\begin{equation}
 \frac{d\sigma}{dx_{\gamma}\,dx_{p}\, dz\, dp_{T}\,dP^2} = \tilde{f}_{\gamma /e}(z,P^2)
 \, q^{\gamma}(x_{\gamma},
Q^2,P^2)\,q^{p}(x_p,Q^2)\,\frac{d\hat{\sigma}}{dp_{T}}
\label{eq:dijeq}
\end{equation}
Here, $p_T$ is the transverse momentum of the jets,  $P^2$ is the
photon virtuality, and $Q^2$ is the relevant energy scale of  the
proccess, taken in this case equal to $p_{T}^2$.
$d\hat{\sigma}/dp_{T}$ represents the hard parton-parton and
parton-photon cross sections \cite{com}.

The functions  $q^{\gamma}(x_{\gamma},Q^2,P^2)$ and
$q^{p}(x_p,Q^2)$ denote the parton
 distribution
 functions for the photon and the proton, respectively. The first one reduces to
  $\delta(1-x_{\gamma})$
-the probability for finding a photon in a photon- for
direct contributions, i.e. those in which the photon participates as such in the hard process.
$\tilde{f}_{\gamma /e}(z,P^2)$ is the unintegrated Weizs\"acker-Williams distribution
\cite{man}
\begin{equation}
\tilde{f}_{\gamma /e}(z,P^2)=\frac{\alpha}{2\pi}\frac{1}{P^2}\frac{1+(1-z)^2}{z}
\end{equation}
which has been shown to be a very good approximation for the distribution of photons
 in the electron,
 provided the photon virtuality is much smaller than the relevant energy scale
\cite{glu}.

\setlength{\unitlength}{1.mm} 
\begin{figure}[t!]
\begin{picture}(150,75)(0,0)
\put(8,12){\mbox{\epsfxsize7.5cm\epsffile{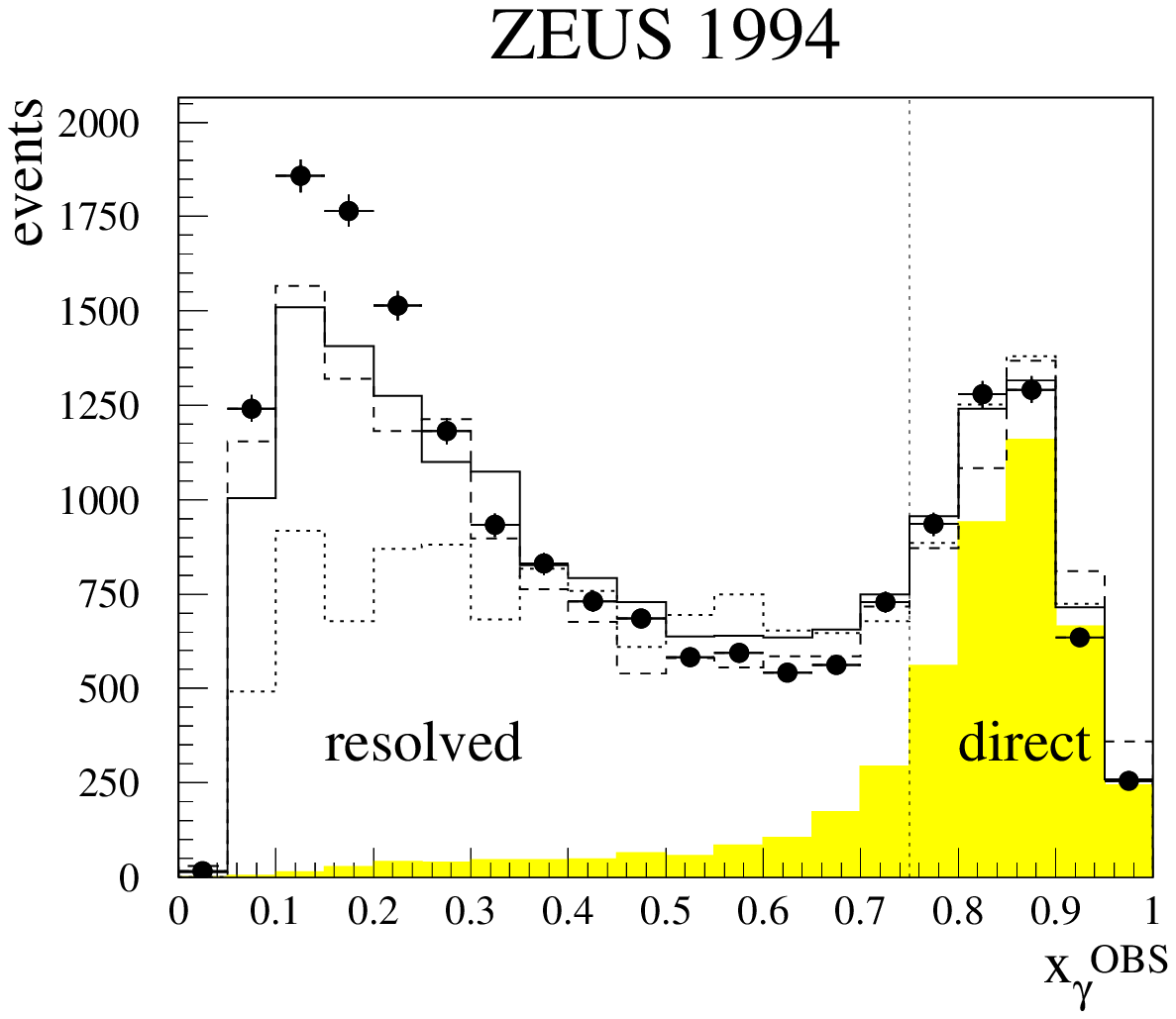}}}
\put(92,13.5){\mbox{\epsfxsize4.5cm\epsffile{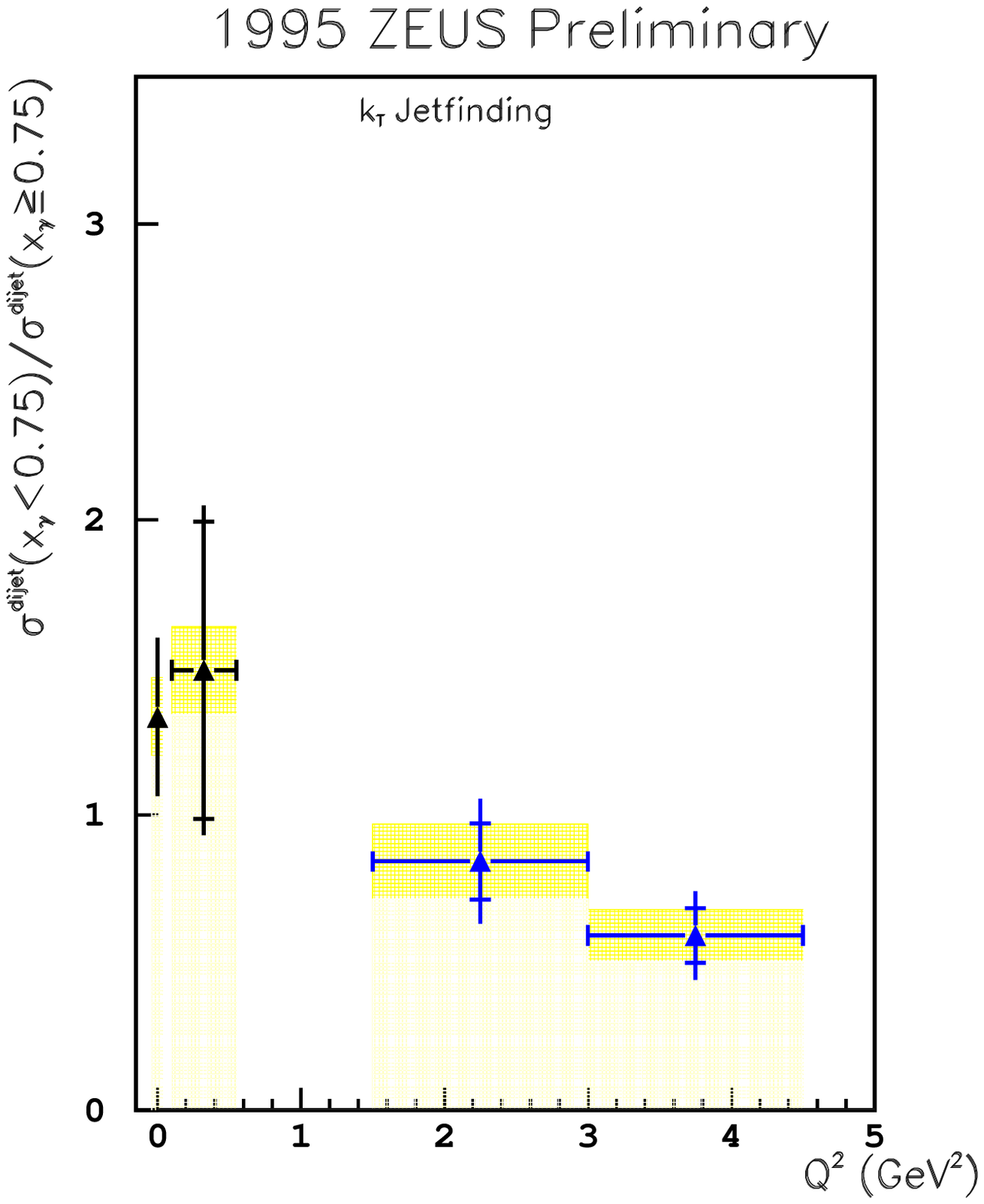}}}
\caption{a) Event distribution measured in dijet photoproduction. b) Ratio of `direct' and `resolved' photon cross sections as a function of the photon virtuality.}
\label{fig:dir/res}
\end{picture}
\end{figure}

The differential cross section in Equation(\ref{eq:dijeq}) can
also be written in terms of the parton pseudorapidities $\eta_1$ and
$\eta_2$, which are constrained by the experimental settings, and
the electron and proton energies $E_e$ and $E_p$ \cite{kram}.
Both pairs of variables are related to the energy fractions by
\begin{eqnarray}
x_p = \frac{p_T}{2 E_p} \left( e^{\eta_1} + e^{\eta_2} \right) \,\,\,\,\,\,\,\, x_\gamma  = \frac{p_T}{2 z E_e} \left( e^{-\eta_1} + e^{-\eta_2} \right) \label{eq:xg}
\end{eqnarray}
Kinematical restrictions constrain $x_\gamma$ to lay in the
interval $[p_T^2 / (x_p z E_e E_p) , x_{\gamma}^{max}]$ , $x_p$ in
$[p_T^2 /(z E_e E_p x_{\gamma}^{max}),1]$ and $z$ in  $[p_T^2/(
E_e E_p),1]$.

\setlength{\unitlength}{1.mm} 
\begin{figure}[b]
\begin{picture}(150,65)(0,0)
\put(45,-8){\mbox{\epsfxsize6.cm\epsffile{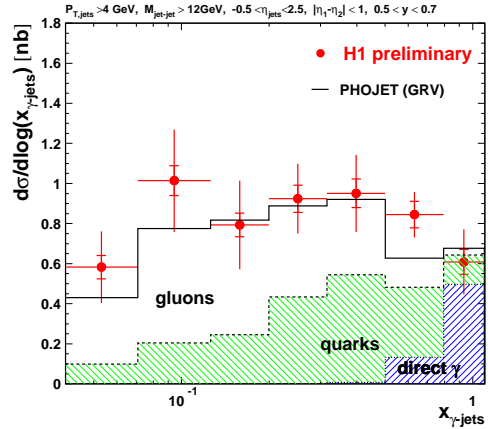}}}
\caption{ Dijet cross sections for $Q^2 < 0.01 GeV^2$ as a function of $x_{\gamma}$.}
\label{fig:q/g}
\end{picture}
\end{figure}

In the lowest order of QCD the `direct' contribution is
characterized by $x_{\gamma}=1$, which means that all the energy
of the photon participates in the hard interaction. Allowing gluon
corrections, `direct' contributions may come from $x_{\gamma} \neq
1$, however one should expect them concentrated arround a peak at
high  $x_{\gamma}$. Conversely, `resolved' contributions should be
more copious al low $x_{\gamma}$ as for ordinary hadrons. In fact,
this trend is shown by the measurements, as can be seen in Figure
\ref{fig:dir/res}a \cite{54levy}. The variable $x_{\gamma}^{obs}$ used there to
approximate $x_{\gamma}$ is the jet level equivalent of Equation
(\ref{eq:xg})
\begin{equation}
x_{\gamma}^{obs}=\frac{p_T^{(1)}\, e^{-\eta_1} +p_T^{(2)}\, e^{-\eta_2} }{2zE_{e}}
\end{equation}
where $p_T^{(i)}$ and $\eta_i$ refer to the transverse energies and
pseudorapidities of the measured jets instead of those of the
partons.   Of course, the relative abundance of `direct' and
`resolved' events depends strongly on the virtuality of the photon
as can be seen in Figure \ref{fig:dir/res}b in accordance to the
expectation about the suppression of the `resolved' component as
the virtuality increases \cite{54levy}.

Finally, it is quite instructive to anlyze the relative abundance
of quark and gluon initiated contributions as a function of
$x_{\gamma}$ in a typical $e^+ p$ dijet cross section. In Figure
\ref{fig:q/g} data obtained by H1 is compared to theoretical
expectations based on a Montecarlo analysis using GRV parton distribution for the photon \cite{glu} and
 for the proton \cite{GRV,9812030}. Notice that at large $x_{\gamma}$ the
`resolved' component is clearly dominated by quarks, however as
$x_{\gamma}$ decreases, the photonic gluons become dominant.

\section*{Colour Singlet (Pomeron?)}

In this section we draw our attention to a much more unfamiliar
object, at least for the younger generation of physicists,
although for some of us it may be an old acquaintance. We are
talking about the {\it pomeron}, or in more modern language, about
the colour singlet object which is exchanged when particles
undergo strong interactions while preserving its nature. The
possibility of studying the partonic structure of this old friend
has been made real by a very recent generation of experiments and
has added an extra quota of excitement in perturbative QCD.

Before we plunge ourselves into the partonic structure of the
pomeron, it would be very helpful to first fetch some of the tools
used in past to sail in those waters.

\subsection*{Peripheral model}

Before the advent of QCD, the presence of structures in the
differential cross-sections was mainly explained by means of
dynamical exchange mechanisms like in the so called {\it
peripheral model}. In this approach it is found that there is a
clear enhancement near the forward direction whenever the crossed
channel of the reaction has the quantum numbers of a known
particle. Then, peripheralism proposes to express the scattering
amplitude as a sum of contributions corresponding to the exchange
of a particle. Namely
\[
F = F_{\pi} + F_V + F_B +...
\]
with
\[
F_i \propto \frac{1}{m_i^2 - t}
\]
where it is clear that for small, physical (negative), values of
$t$, there are peaks in the corresponding cross-section in
s-channel, proportional to $|F|^2$. The indices $\pi, V, ...$
represent pions, vector bosons, etc., and their masses hierarchize
each contribution. The explanation of peripheral
events cannot be found in QCD, since the exchanged particle is not
a gauge boson that can be treated perturbatively. In fact the
particles involved are composite objects such mesons or baryons.

When the exchanged particle has a spin $J$, the previous
expression for the amplitude has to be replaced by
\[
F_i \propto \frac{P_J(\cos\theta_t)}{m_i^2 - t}
\]
and, if one remembers that the scattering angle in the $t$-channel
reads
\[
\cos \theta_t = 1 + \frac{2\,s}{t - 4\,m^2}
\]
it is found that the numerator of $F_i$ behaves asymptotically as
$s^J$ for fixed $t$. This is not tenable because the Froissart
bound is not satisfied for $J > 1$. Moreover, that expression is
purely real and does not satisfy the principle of analyticity.

There is a way out of the above mentioned difficulties, while
maintaining the spirit of an exchange mechanism, now of an entire
family of related particles. This proposal is based on the Regge
poles ideas. These ideas emerge naturally in potential scattering
theory through the analyticity properties of the amplitude. The
Regge pole model \cite{REG} is based then, upon the assumption of
the following behaviour for the scattering amplitude \beq \lim_{s
\rightarrow \infty\,,\,fixed\,t}\,F(s,t) \simeq \sum_i
\beta_i(t)\,\frac{1 + \xi_i\,e^{-i\,\pi\,\alpha_i(t)}} {\sin
[\pi\,\alpha_i(t)]}\,\left(\frac{s}{s_0}\right)^{\alpha_i(t)} \eeq
here $\xi_i = \pm 1$ is called the signature of the corresponding
Regge pole. The functions $\alpha_i(t)$ are called {\it Regge
trajectories} and $\beta_i(t)$ residues of the poles that in
general factorize between the initial and the final channels. Both
functions $\alpha_i(t)$ and $\beta_i(t)$ are analytic functions of
$t$.

The high energy model so introduced is understood as an exchange
model where the Regge trajectories $\alpha_i(t)$ are such that
they pass through the spin values of  the particles, or better the
resonances, each time positive $t$ takes the values of their mass
squared. In this way, the collective effect of the exchange of all
members of the family is taken into account. Clearly, the
Froissart bound (total cross-sections cannot increase faster than
$\ln^2\,s$) is satisfied as soon as
\[
\alpha_i (t \le 0) \le 1
\]
in the space-like region of negative $t$, relevant for the
sacttering process.

 Let us take as an example the case of
pion-nucleon charge-exchange
\[
\pi^-\,\,p \rightarrow \pi^0\,\,n
\]
that has always been the paradigmatic case of reggeology. The
$t$-channel of this reaction is
\[
\pi^-\,\,\pi^0 \rightarrow \bar{p}\,\,n
\]
having the quantum numbers; $Q = 1$, $I = 1$, $Y=0$. $G= +1$ that
correspond to the well known meson resonances $\rho\,(770\,MeV)$
of spin 1 and $g\,(1680\,MeV)$ of spin 3. This sequence of
resonances defines precisely the meson Regge trajectory of isospin
$1$, G-parity $+1$ and positive signature that is noted
$\alpha_{\rho}$. The continuation of this trajectory to negative
$t$ values, takes us into the physical region corresponding to the
$s$-channel charge-exchange reaction. The parameters of a standard
linear trajectory are the intercept at $t = 0$ and the slope. In
the present case, as it is shown in Figure \ref{fig:RHO}, they are

\setlength{\unitlength}{1.mm} 
\begin{figure}[hbt]
\begin{picture}(150,50)(0,0)
\put(15,-95){\mbox{\epsfxsize12.cm\epsffile{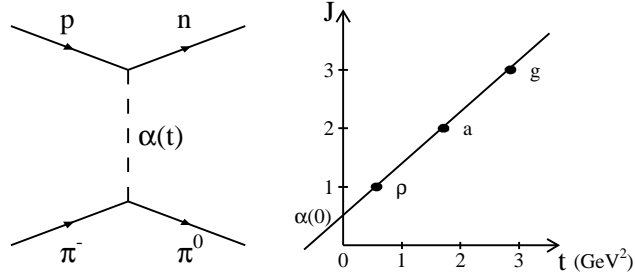}}}
\vspace{10pt}
\caption{$\rho$-Regge trajectory and $\pi^-\,\,p \rightarrow \pi^0\,\,n$ data }
\label{fig:RHO}
\end{picture}
\end{figure}

\[
\alpha_{\rho} (0) \simeq 0.55\,\,\,\,\,\alpha^{\prime} \simeq
0.9\,GeV^{-2}
\]
respectively. Consequently, the differential cross section has the
behaviour
\[
\left. \frac{d \sigma}{d t} \right|_{t = 0} \simeq
s^{2\,(\alpha_{\rho}(0) - 1)} \sim s^{-1}
\]

But what happens if the crossed channel of the reaction, and
correspondingly the eventual exchanged object, has the quantum
numbers of the vacuum? The answer arrives in the following
paragraphs.

\subsection*{Diffraction}

Landau and his collaborators, in the fifties, introduced the term
diffraction in high energy physics used in complete analogy with
the well known phenomenon in Optics that occurs when light
interacts with obstacles or holes whose dimensions are of the
order of the electromagnetic radiation wavelength. The interaction
of a hadron could be think as the absorption of its wave function
caused by the different channels open at high energy and
consequently the name diffraction seems to be intuitive. For a
recent review of this topic see \cite{PREDA}.

In the Fraunhofer limit, namely when the product of the wave
number times the area $\Sigma$ of the obstacle is of the order of
the observation distance, the energy distribution at the
observation point, given by the Kirchoff formula, can be written
as
\[
T(x,y,z) \approx \frac{k}{2\,\pi\,i}\,\frac{e^{i\,k\,r_0}}{r_0}
\int_{\Sigma} d^2b\,S(\vec{b})\,e^{i\,\vec{q}\cdot\vec{b}}
\]
where $\vec{b}$ stands for the impact parameter, $r_0$ is the
position of $\Sigma$, $|\vec{q}| = k\,\sin \theta$ is the 2-D
momentum transfer and the scattering matrix is expressed as
\[
S(\vec{b}) = 1 - \Gamma(\vec{b})
\]
in terms of the so called profile function $\Gamma(\vec{b})$. From
the expression for $T$, one can immediately obtain the scattering
amplitude \beq f(\vec{q}) = \frac{i\,k}{2\,\pi} \int
d^2b\,\Gamma(\vec{b})\,e^{i\,\vec{q}\cdot\vec{b}} \eeq i.e.: given
as the 2-D Fourier transform of the profile function.

When the function $\Gamma$ is spherically symmetric, the integral
becomes a Bessel transform, namely
\[
f(\vec{q}) = i\,k \int^{\infty}_0
b\,db\,\Gamma(\vec{b})\,J_0(q\,b)
\]
meaning that if the profile function is merely a disk of radius
$R$, the amplitude reduces to the black-disk form
\[
f(\vec{q}) = i\,k\,R^2\frac{J_1(q\,R)}{q\,R}
\]
The prediction coming from this very simple model is a series of
diffractive maxima and minima, entirely similar to the case in
Optics, and have been clearly observed in several experiments

In the specific field of particle physics, diffraction is said to
be the dominant process of scattering at high energy if no quantum
numbers are exchanged between the colliding particles. In other
words, diffraction dominates asymptotically as soon as the
particles in the final state have the same quantum numbers of the
incident ones. This sort of definition of diffraction clearly
includes, for the two body scattering, three cases: elastic
scattering, single diffraction and double diffraction. In the
first process the outgoing particles are exactly the same as the
incident ones. In the second case, one incident particles
goes out unmodified while the second one gives rise to a
resonance, or to a bunch of final particles, with total quantum
numbers coincident with its own ones. Finally, when double
diffraction occurs, each incident particle gives rise to a
resonance, or to a bunch of final particles, with the same quantum
numbers of the initial ones.

\subsubsection*{Pomeron}

\ni It is possible to discuss diffraction from the viewpoint of an
exchange model, in particular within the framework of reggeology.
Remember that we have found precisely that an enhancement in the
differential cross section can be expected whenever the quantum
numbers of the crossed channel correspond to an existing particle.
However, in the elastic case at high energies, all the hadronic
systems show the diffraction peak while the exchange of a given
particle gives rise to different contributions in different
systems. Moreover, via the optical theorem, the forward elastic
amplitude is connected to the total cross section and this fact
cannot be understood in terms of only one exchange process.
Nevertheless, the diffraction peak can be interpreted in terms of
a special trajectory that summarizes all the diffractive
contributions and has the vacuum quantum numbers: the {\it
pomeron}. Its contribution to the amplitude is
\beq F_{dif}(s,t) =
- \beta_{\pp}(t)\,\frac{1 + e^{-i\,\pi\,\alpha_{\pp}(t)}}{\sin
[\pi\,\alpha_{\pp}(t)]}\, \left( \frac{s}{s_0} \right) ^{\alpha_{\pp}(t)}
\eeq Consequently, the asymptotic behaviour of the total cross
section (via optical theorem) is given by
\[
\sigma_{tot} \simeq \frac{\beta_{\pp}(0)}{s_0}
\]
if the intercept of the pomeron trajectory is the maximum value
allowed by the Froissart bound, namely $\alpha_{\pp}(0) =1$. This fact
implies that the total cross section in this approximation behaves
asymptotically as a constant. It is clear that logarithmic
corrections are always allowed. Moreover, they are necessary in
order to cope with experimental data.

The pomeron concept answer the question at the end of the previous section using
the reggeon jargon: a diffraction process is dominated by the
exchange of a {\it pomeron}, that in fact means exchange of no
quantum numbers.

\begin{figure}[t!] 
\centerline{\epsfig{file=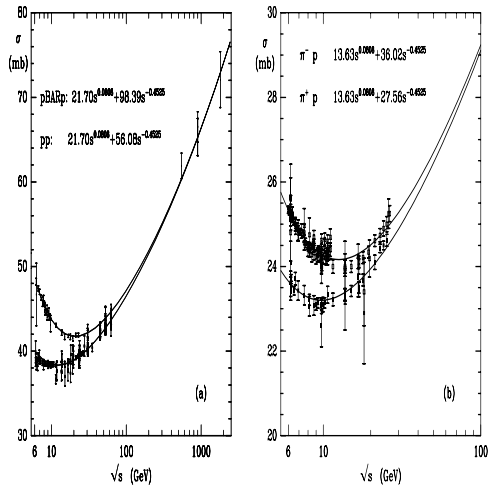,height=2.5in,width=2.5in}}
\vspace{10pt} \caption{Donnachie and Landshoff parametrization of
total cross sections} \label{fig:DAL}
\end{figure}

It is interesting to mention that Donnachie and Landshoff
\cite{DAL} were able to describe all available total cross section
data on $\bar{p}p$, $pp$, $\pi p$ and $Kp$ by a very simple
Regge-Pomeron inspired parametization of the form
\[
\sigma_{tot} = X\,s^{0.0808} + Y\, s^{-0.4525}
\]
where $X$ and $Y$ are reaction dependent parameters. Clearly the
first exponent of the center of mass energy squared $s$ correspond
to a pomeron intercept of $\alpha_{\pp}(0) = 1.0808$, while the second
exponent comes from a typical Regge intercept $\alpha_{\rr}(0) =
0.5475$. The successful fit is shown in Figure \ref{fig:DAL}.

Trying to connect the pomeron with the language and understanding
of QCD, one can imagine it as a colour singlet combination of
partons such as the simplest picture of a pair of gluons
proposed by Low and Nussinov \cite{LAN}. Two-gluon exchange is
compatible with all soft phenomenology except that this simple
model gives a constant, not a rising, total cross-section. Clearly
in order to analyze the parton content of pomeron, a hard or
high-momentum transfer interaction analogous to the one used to
find the quark-gluon content of the proton is necessary. This kind
of processes is called hard diffraction making reference to the
hard scale that ultimately allows the pertubative description.

Nowadays, people is trying to go beyond in the understanding of
pomeron properties following different theoretical
approaches\cite{HARDIFF}, although there remain several open
questions. The main point to be solved is related to the precise
relation between hard and soft diffraction. As we said, to study
the pomeron parton content, small distances and high-momentum
transfer are necessary, while the natural environment of the soft
pomeron is at large distances and low-momentum transfer. In other
words, the clear notion of a pomeron is still missing.


Most of the recent renewal of interest in diffraction 
was triggered mainly by a theoretical proposal of Bjorken
\cite{BJD}, who pointed out a new signature of diffraction related
to the presence of large rapidity gaps. These rapidity gaps were
found both at HERA and at the Tevatron.

 The presence of diffractive processes at HERA should be
expected since the photon behaves like a hadron in several
circumstances. In fact, about $40\,\%$ of the photoproduction
events are of diffractive character. In this kind of processes,
the pomeron is exchanged and the incident proton remains as a
proton or is diffractively dissociated in a state with the same
quantum numbers. Consequently, as it will be discussed below,
there appears a large rapidity gap between the proton, or the
diffractive system, and the hadrons coming from the system into
which the photon was diffracted. The quite unexpected fact
observed at HERA was the presence of large rapidity gap events
also in the DIS domain \cite{DISD}. The observation that also a highly
virtual photon can participate in a diffractive process, opened
the road to the analysis of the pomeron structure. In fact, if $Q^2$, the
virtuality of the photon,  is larger than say $4\,GeV$, the
scattering $\gamma^{\star}$-pomeron can be treated perturbatively.
The impact of this DIS diffractive data is also evident because
one is facing in a DIS process characterized by a large $Q^2$
scale, diffractive properties which were been expected at a soft
scale.

\subsubsection*{Large rapidity gaps}

As it was predicted by Bjorken \cite{BJD}, the most evident signal
for diffractive physics at high energy, is the presence of large
rapidity gaps. Just to get some insight on this jargon, remember
that a single inclusive diffractive reaction, noted as
\[
A(k_A) + B(k_B) \rightarrow A' (k^{\prime}_A) + X(k_X)
\]
implies no exchange of quantum numbers different from those of
vacuum. In this inclusive case, at variance with the elastic
reaction, besides the two variables ($k$ and $\theta$ or $s$ and
$t$) a third variable is needed to describe the process. Generally
it is used \beq M_X^2 = (k_A + k_B - k^{\prime}_A)^2 = E_X^2 -
\vec{k}^2_X
 \eeq
 or, alternatively, the Feynman $x_F$ defined as
 \beq
 x_F
\equiv \frac{|k^{\prime}_{A\,long}|}{k^{\prime}_{A\,long}}\approx1
- \frac{M^2_X}{s}
 \eeq
 Another useful variable is rapidity,
defined by
 \beq
  y = \ln \left[\frac{E^{\prime}_A +
k^{\prime}_{A\,long}}{E^{\prime}_A - k^{\prime}_{A\,long}}\right]
\eeq or equivalently, the pseudorapidity which is its limit for
large energy
 \beq
\eta = - \ln\left(\tan\,\frac{\theta_{A'}}{2} \right)
 \eeq

In a typical DIS event, the struck parton of the proton, emerges
forming an angle $\theta_{A'}$ with the proton remnant
direction, as it is shown in Figure \ref{fig:REM}.

\setlength{\unitlength}{1.mm} 
\begin{figure}[hbt]
\begin{picture}(150,55)(0,0)
\put(30,-62){\mbox{\epsfxsize9.0cm\epsffile{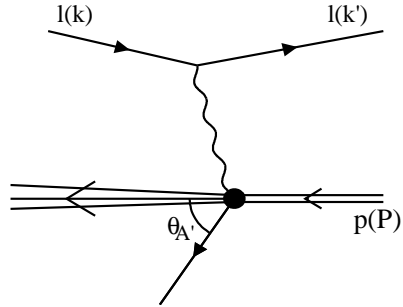}}} \caption{DIS
event diagramm} \label{fig:REM}
\end{picture}
\end{figure}

This angle can be expressed in terms of the difference in
total pseudorapidity $\eta$ between these directions, namely
\[
\triangle \eta = \eta_{remnant} - \eta_{parton}
\]
If the center of mass energy squared of the system is $s$, the
pseudarapidity interval covered is
\[
 \triangle \eta \sim \ln \left( \frac{s}{m_p^2} \right)
\]
Consequently,
\[
\triangle \eta \sim \ln \left( \frac{W}{m_p} \right) - \ln \left(
\frac{x\,W}{m_p} \right) \sim \ln \left( \frac{1}{x} \right)
\]
where the first term correspond to the pseudorapidity covered by
the $\gamma^{\star}-p$ system and the second to that  covered by
the $\gamma^{\star}-quark$ system respectively. Clearly $x$ is the
Bjorken variable that measures the amount of momentum of the
proton carried by the quark.

The confinement property of QCD can be rephrased by saying that
the struck quark and the proton remnant are connected via a colour
string in order to end with colourless hadrons. For this reason, the mentioned pseudorapidity gap
$\triangle \eta$ is filled with particles during the hadronization
process. Moreover, when the value of $x$ decreases, the average
hadron multiplicity increases, making less likely the visibility
of any rapidity gap in the DIS event. In principle, the rapidity
gap between the remnant proton, and the struck quark jet is then
exponentially suppressed.

After this kinematical preface, we briefly discuss the so called
{\it leading particle effect}, characteristic of high energy
hadronic interactions. Around $10 \%$ of the inclusive hadron
scattering events present a Lab system configuration where the
incident particle flies apart essentially unscattered in almost
the forward direction leaving behind a stream of slow moving
produced particles. The experimentally determined cross section
looks entirely similar to the elastic case. Namely, between the
initial particle and the final fast one, there is no change of
quantum numbers and the reaction is clearly of diffractive type.
There is only a minor loss of momentum to produce the slow
particles. This effect requires the fast particle to be exactly
the same as the incident one.

When the leading particle, the one we called above $A'$, is
produced diffractively, and because the process has this
characteristic, one has to expect the remaining cluster $X$ to be
produced at the opposite end of the rapidity spectrum. That is
usually expressed by saying that the reaction presents a large
rapidity gap from $A'$. Diffraction definition makes
evident this concept. In fact, as $A'$ has exactly the same
quantum numbers as $A$, no quantum numbers at all can be exchanged
until $X$ is produced. If a particle were produced in between, it
would mean that some quantum number has been exchanged and the
process is no longer diffractive. Consequently, in this large
rapidity gaps events, as no particle is produced between
$A'$ and $X$, they are clearly separated in rapidity.

Large rapidity gap events can be generated in soft processes as
elastic and single diffractive hadron-hadron scattering, where the
momentum transfer $t$ between the scattered particles is small,
and the center of mass energy $s$ large, i.e., peripheral
scattering.

It should be stressed that nowadays, diffraction is considered
synonym of large rapidity gap processes. In other words, the
evidence of diffractive hadronic events comes mainly from large
rapidity gaps as have been observed in D0 and CDF Collaborations
at the Tevatron \cite{LRGE}.

On the other hand, at HERA, the electron-proton collider,
diffraction has also been extensively observed using both the
rapidity gap criteria for diffraction and also the detection of
final state protons. In semi-inclusive DIS, whenever the detected
final hadron coincides with the incident one, one is dealing with
diffraction. In order to follow the outgoing proton a Leading
Proton Spectrometer (LPS) was designed.

\subsection*{Hard Diffraction at HERA}

Before discussing the interesting HERA DIS diffractive events, let
us consider the particular semi-inclusive processes where the
detected final hadron exactly coincides with the initial one,
namely
\[
\ell(k_{\ell}) + N(k_N) \rightarrow \ell(k^{\prime}_{\ell}) +
N(k_N^{\prime}) + X(k_X)
\]
In this case the scattering diagram looks like the one in Figure
\ref{fig:LRG}

\setlength{\unitlength}{1.mm} 
\begin{figure}[hbt]
\begin{picture}(150,56)(0,0)
\put(25,-60){\mbox{\epsfxsize9.0cm\epsffile{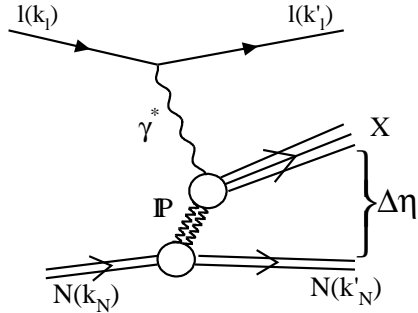}}}
\caption{Diffractive DIS event diagramm} \label{fig:LRG}
\end{picture}
\end{figure}

The process is clearly diffractive since no quantum numbers are
exchanged between the $\gamma^{\star}$ and the nucleon $N$. As a
consequence, one has for the remnant $X$ the quantum numbers
$J_X^{PC} = 1^{--}$ exactly equal to those of the incoming photon.
Notice by the way that the production of any vector boson, the
explicit replacement of $X$ by any $V$ in the reaction above, is a
diffractive process because the $V$ meson quantum numbers are
precisely those of $\gamma^{\star}$.

The kinematics related to Figure \ref{fig:LRG} needs further
variables to be introduced. They are usually defined  as \beq t  =
(k_N - k_N^{\prime})^2  \,\,\,,\,\,\, M_X^2  =   (k_N -
k_N^{\prime} + q)^2 \,\,\,,\,\,\,  M_Y^2  =  k_N^{\prime\,^2} \eeq
\beq
 x_{\pp}  =  \frac{(k_N - k_N^{\prime})\cdot q}{k_N \cdot k_N^{\prime}}
 = \frac{M_X^2 + Q^2 -t}{W^2 + Q^2 - m_p^2} \,\,\,,\,\,\,
\beta  =  \frac{Q^2}{M_X^2 + Q^2 - t} = \frac{x}{x_{\pp}}
\eeq
Notice that frequently $\xi$ is used instead of $x_{\pp}$ and that
$x_F = 1 - x_{\pp}$. Clearly, the range of values of $x, x_{\pp},
x_F$ and $\beta$ is $(0,1)$. Moreover, $\beta$ could be understood
as the momentum fraction carried by the parton directly coupled to
$\gamma^{\star}$.

In the context of diffractive DIS, it is customary to define {\it
diffractive structure functions}, in analogy with the ordinary
ones, through the expression of the corresponding cross section
\begin{eqnarray}
\frac{d^4\sigma (\ell\,N \rightarrow
\ell\,X\,Y)}{dx\,dQ^2\,dx_{\pp}\,dt}  =
\frac{4\,\pi\,\alpha^2_{em}}{x\,Q^4}  \left[
\left(1 - y +\frac{y^2}{2} \right)\,F_2^{D(4)} -
\frac{y^2}{2}\,F_L^{D(4)}\right]
\end{eqnarray}
where
\[
F_2^{D(4)} = F_T^{D(4)} + F_L^{D(4)}
\]
It is also of interest the $t$-integrated diffractive structure
function defined en each case by \beq F_i^{D(3)}(x_{\pp}, \beta,
Q^2)=\int_{|t|_{min}}^{|t|_{max}} d|t|\, F_i^{D(4)} (x_{\pp},
\beta, Q^2, t) \label{eq:FD}
 \eeq
 with $i=2, T, L$. The superindices $3$ and $4$ refer to
the number of variables in the structure functions. In the
integral $|t|_{min}$ is the lower kinematic limit of $|t|$ and
$|t|_{max}$ has to be specified in each case.

In the most naive Regge inspired approach, the diffractive
structure function is assumed to be given by the product of the
probability $f_{\pp /p}(x_{\pp},t)$ to find a pomeron in the
incoming proton, which only depends on the variables $x_{\pp}$ and
$t$, and a {\it pomeron structure function}
$F_2^{\pp}(\beta,Q^2)$, given by parton densities which are
assumed to behave according to Altarelli-Parisi evolution
equations and factorize as ordinary parton distributions
\cite{KUN}.

 In recent years different theoretical approaches have been
proposed for the description of hard diffraction and specifically
for the diffractive structure functions. See for example
\cite{HARDIFF} and references therein.

Some of these approaches  modelize the diffractive interaction in
the proton rest frame as an exchange of two gluons between the
proton and the hadronic system into which the photon has evolved
as depicted by the upper line diagramas in Figure \ref{fig:gl}.
Others, also in the proton rest frame, evaluate within a
semiclassical analysis the effect of the colour field of the proton
into the photon hadronic system as in the lower line diagrams of
Figure \ref{fig:gl}. The main difference between the above
mentioned models is that in the former only two gluons are
exchanged while in the latter a multiple exchange of soft gluons
is assumed \cite{HARDIFF}.

\setlength{\unitlength}{1.mm} 
\begin{figure}[t!]
\begin{picture}(150,70)(0,0)
\put(30,10){\mbox{\epsfxsize4.0cm\epsffile{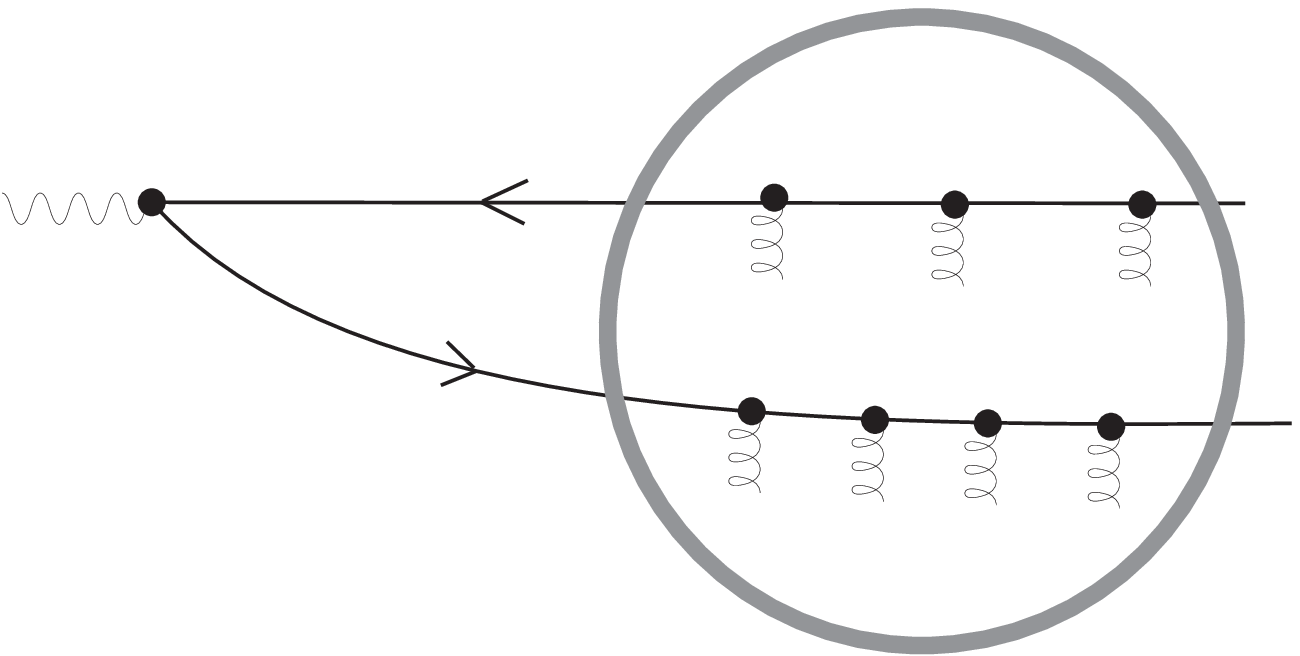}}}
\put(80,10){\mbox{\epsfxsize4.0cm\epsffile{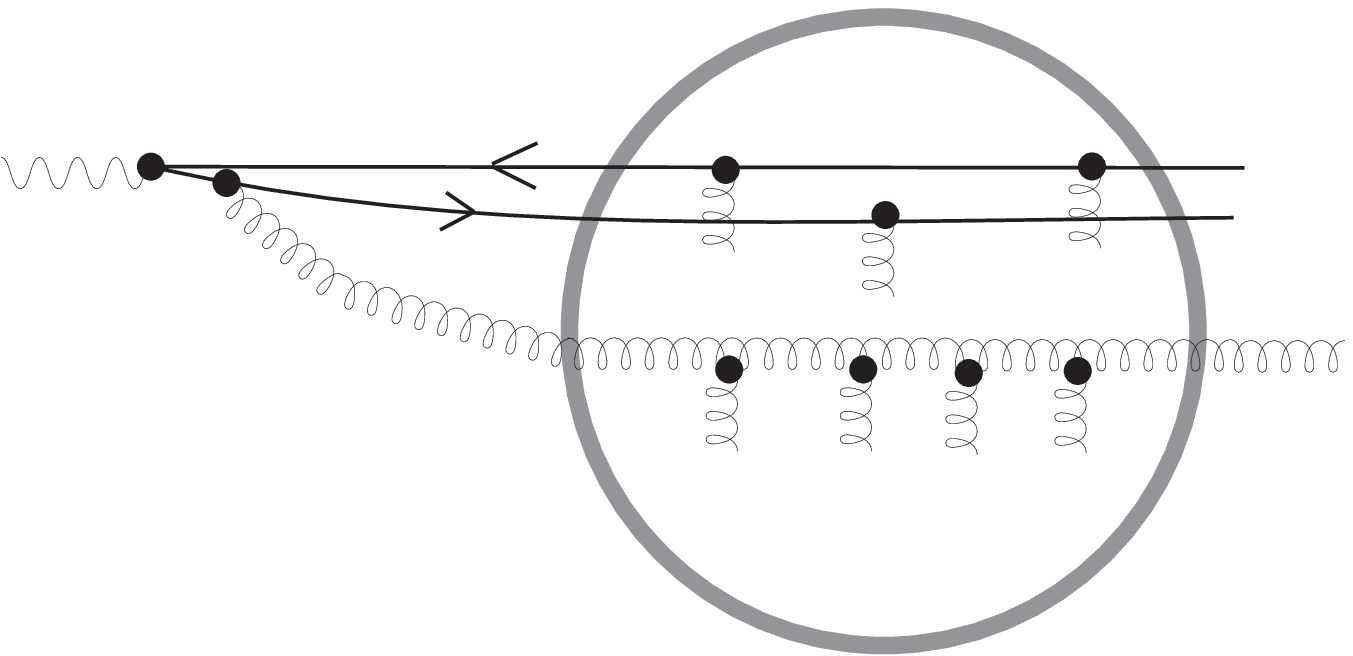}}}
\put(30,40){\mbox{\epsfxsize4.0cm\epsffile{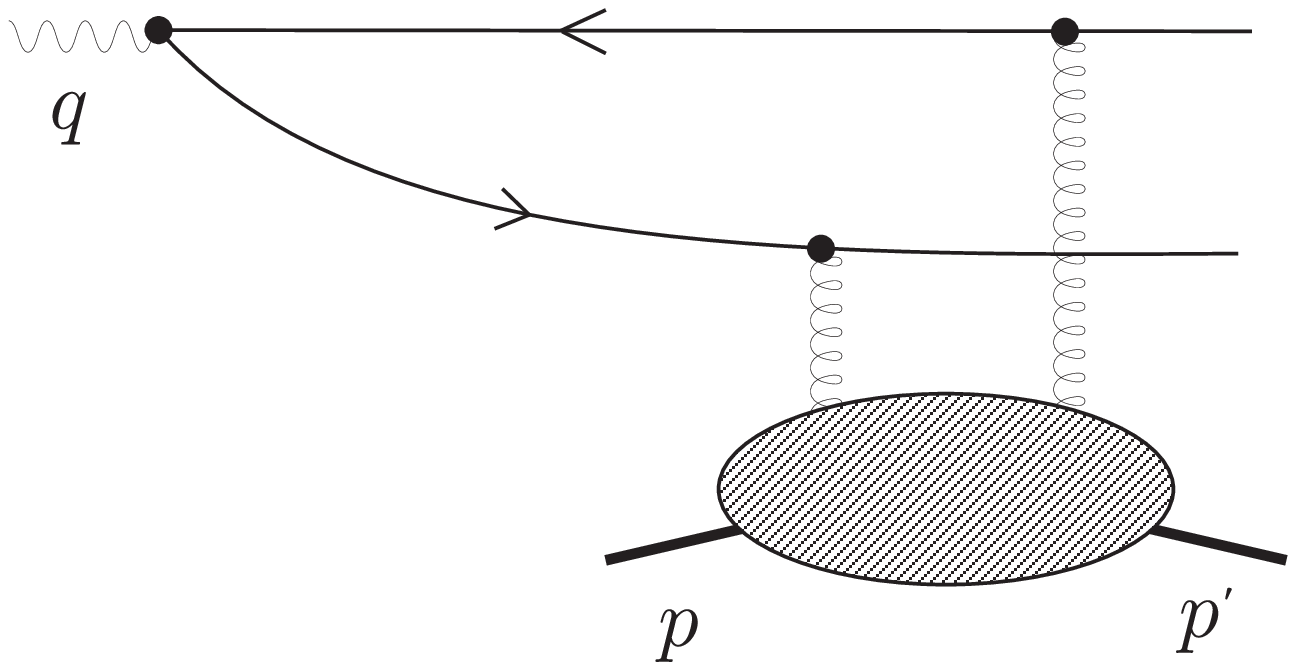}}}
\put(80,40){\mbox{\epsfxsize4.0cm\epsffile{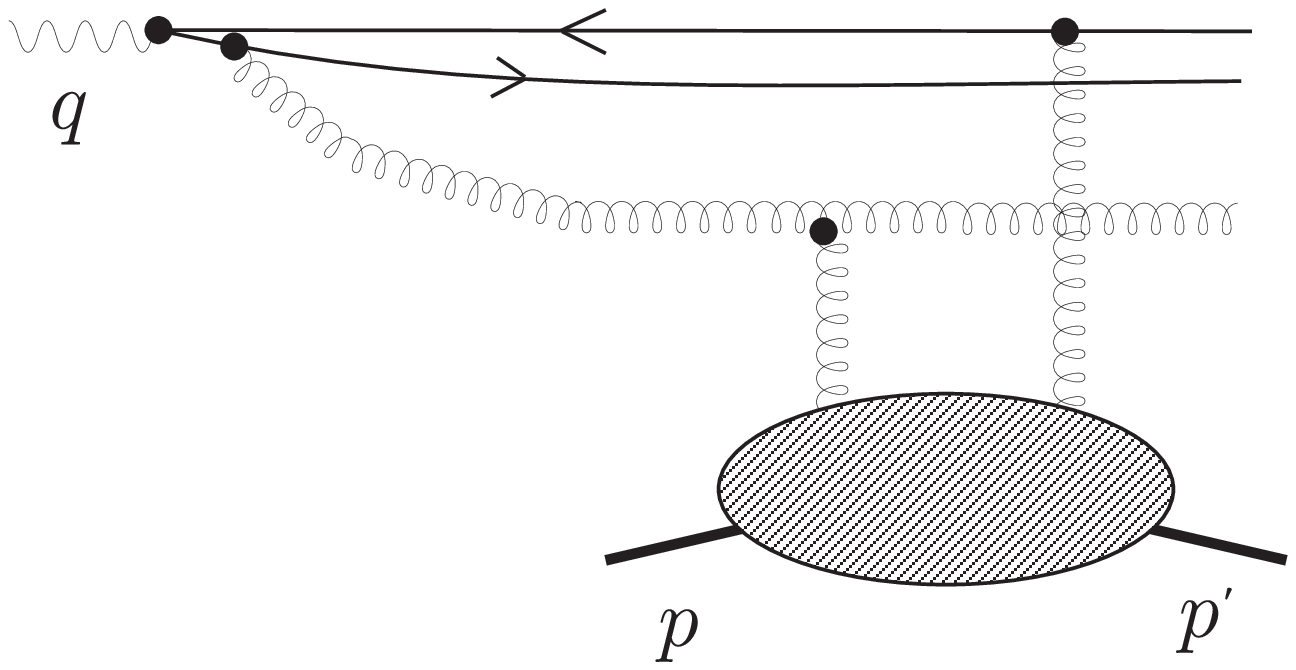}}}
\caption{Different approaches for diffractive DIS} \label{fig:gl}
\end{picture}
\end{figure}

\setlength{\unitlength}{1.mm} 
\begin{figure}[b!]
\begin{picture}(150,50)(0,0)
\put(10,20){\mbox{\epsfxsize5.0cm\epsffile{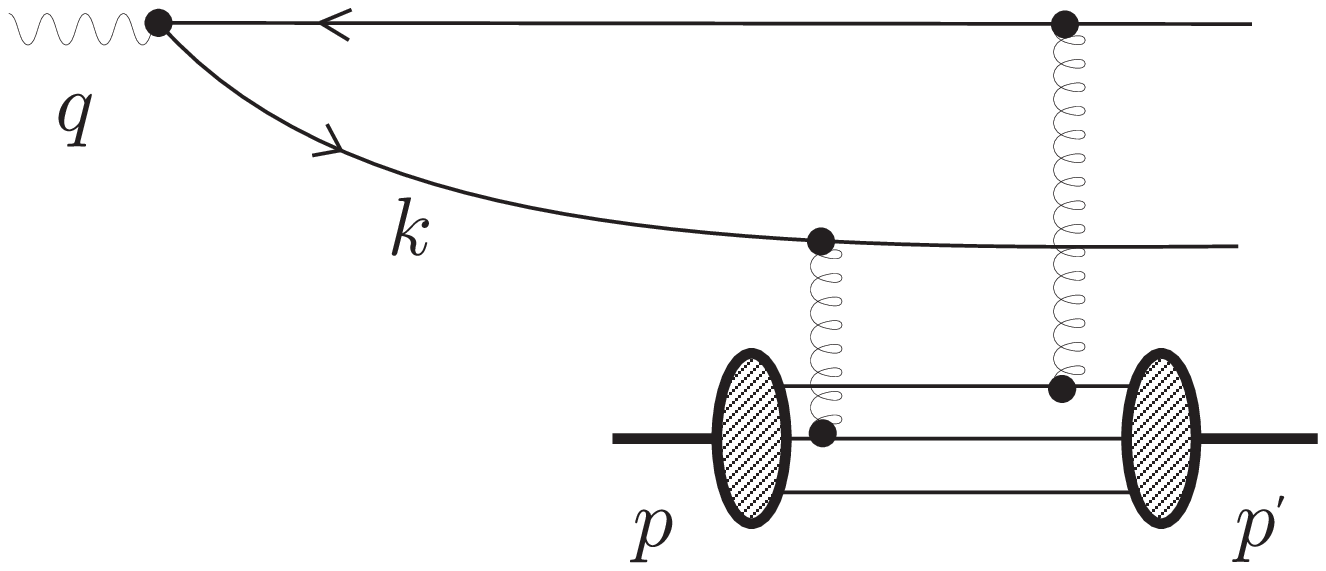}}}
\put(90,20){\mbox{\epsfxsize5.0cm\epsffile{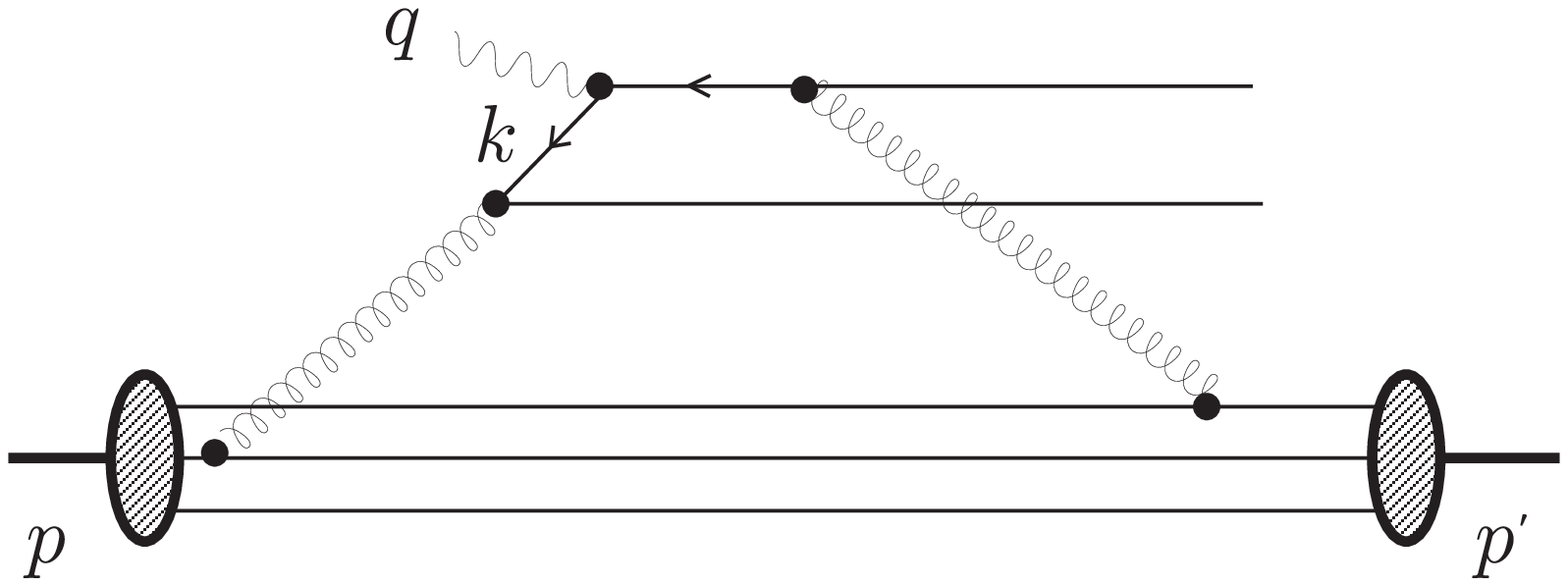}}}
\put(70,30){\mbox{{\LARGE $\longleftrightarrow$}}}
\caption{Correspondence between the two gluon exchange models and
the factorization approach.} \label{fig:trans}
\end{picture}
\end{figure}

A third strategy consists in dealing with the diffractive
interaction as a special kinematical limit of a more general
semi-inclusive process, i.e. regarding the diffractive structure
function just as the low $x_{\pp}$ limit of the fracture function
of protons into protons \cite{VEN}.  Doing this, the whole
perturbative techniques can be rigurously applied without need to
make additional assumptions in a program similar to what has been
done for structure and fragmentation functions \cite{ph}. Another
advantage of this last
 strategy is that in this framework the large $x_{\pp}$ behaviour of the
diffractive structure function can be explored with leading proton
production experiments. The fracture function approach can be in some sense related
to the two gluon exchange model by a frame transformation  as it is sketched in Figure \ref{fig:trans}.

A remarkable thing to notice regarding these kinds of models for
diffraction and also the factorization approach, is that although
they seem to differ in rather strong assumptions, they all give a
reasonably good account of the data suggesting a large gluon
content in the diffractive structure
 function with gluons concentrated at high $\beta$.

\setlength{\unitlength}{1.mm} 
\begin{figure}[t!]
\begin{picture}(150,100)(0,0)
\put(30,-10){\mbox{\epsfxsize9.0cm\epsffile{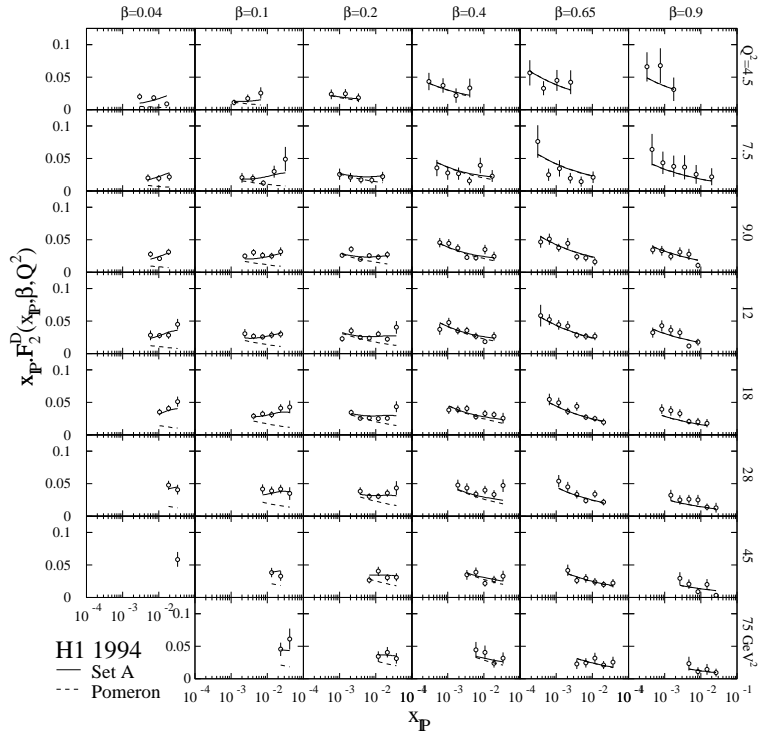}}}
\caption{Diffractive DIS data.} \label{fig:F2D}
\end{picture}
\end{figure}

\setlength{\unitlength}{1.mm} 
\begin{figure}[b!]
\begin{picture}(150,85)(0,0)
\put(40,-5){\mbox{\epsfxsize7.5cm\epsffile{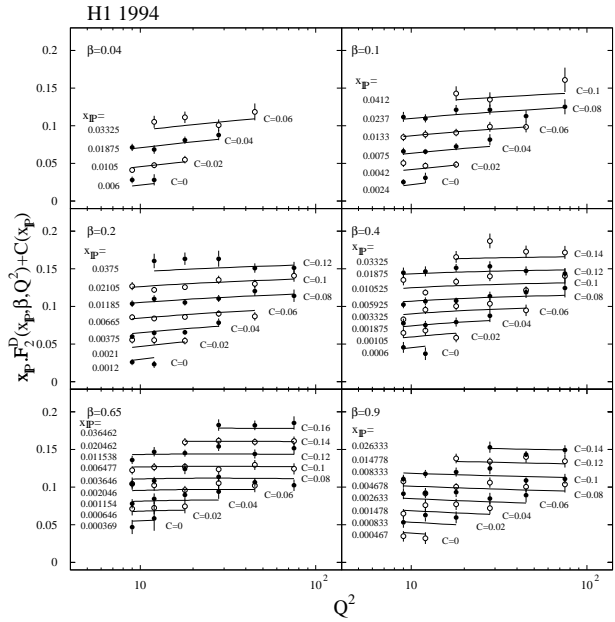}}}
\caption{Scale dependence of $F_2(x_{\pp},\beta,Q^2)$.}
\label{fig:F2DQ2}
\end{picture}
\end{figure}

In Figure \ref{fig:F2D}
 measurements of the diffractive structure function $F_2(x_{\pp},\beta,Q^2)$
 as a function of $x_{\pp}$ obtained by H1 at HERA are compared with the fit
 coming from a factorization approach \cite{ph}.As for ordinary structure functions, QCD predicts the scale
dependence of the diffractive structure function. Figure
\ref{fig:F2DQ2} shows the agreement between the data and the
behaviour expected in the factorization approach.

Having established a rigorous and precise description of
diffractive DIS we can go back to the most naive Regge inspired
approach and see which of their hypothesis or assumptions may
survive. The first thing to notice is that although in principle
diffractive parton distributions or fracture functions obey non
homogeneus AP evolution equations, in the low $x_{\pp}$ limit the
non homogeneus contributions are numerically negligible so the
standard assumption is a very good approximation. Instead, what
fails, at least in the unrestricted kinematical range accesed by
HERA, is the flux factorization hypothesis, i.e. the posibility to
factorize the $x_{\pp}$-dependence of the structure function as a
simple power of $x_{\pp}$. Global analysis of the data shows not
only deviations form this behaviour, suggesting the admixture of
Regge exchanges, but also a $\beta$-dependent level of admixture  \cite{ph}.

HERA also measures other diffractive procceses like diffractive
 production
of vector mesons and diffractive photoproduction of jets
\cite{summa}. This last kind of processes, depicted in Figure
\ref{fig:photog},
is particularly exciting as it represents a combined test of our
 knowledge of the parton structure of the photon and the colour singlet.
>From the theoretical point of view it is also very important
because the resolved photon contribution to the process is in fact
a hadron-hadron diffractive process for which hard QCD
factorization may be broken.
\setlength{\unitlength}{1.mm} 
\begin{figure}[b!]
\begin{picture}(150,55)(0,0)
\put(60,-60){\mbox{\epsfxsize9.0cm\epsffile{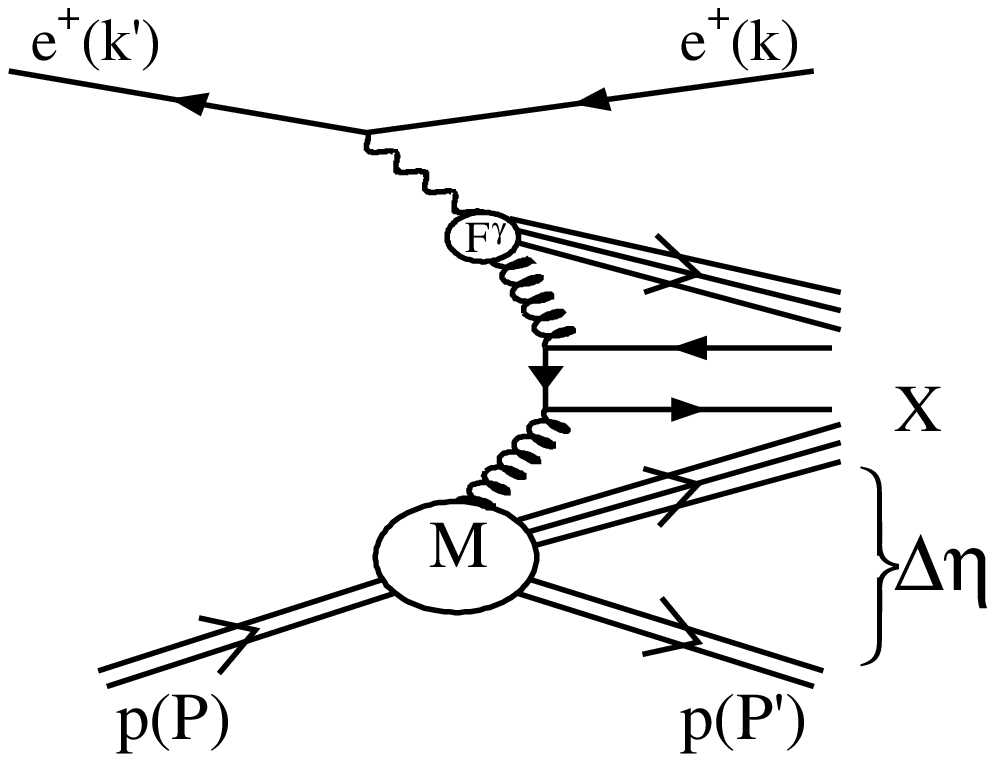}}}
\put(-5,-60){\mbox{\epsfxsize9.0cm\epsffile{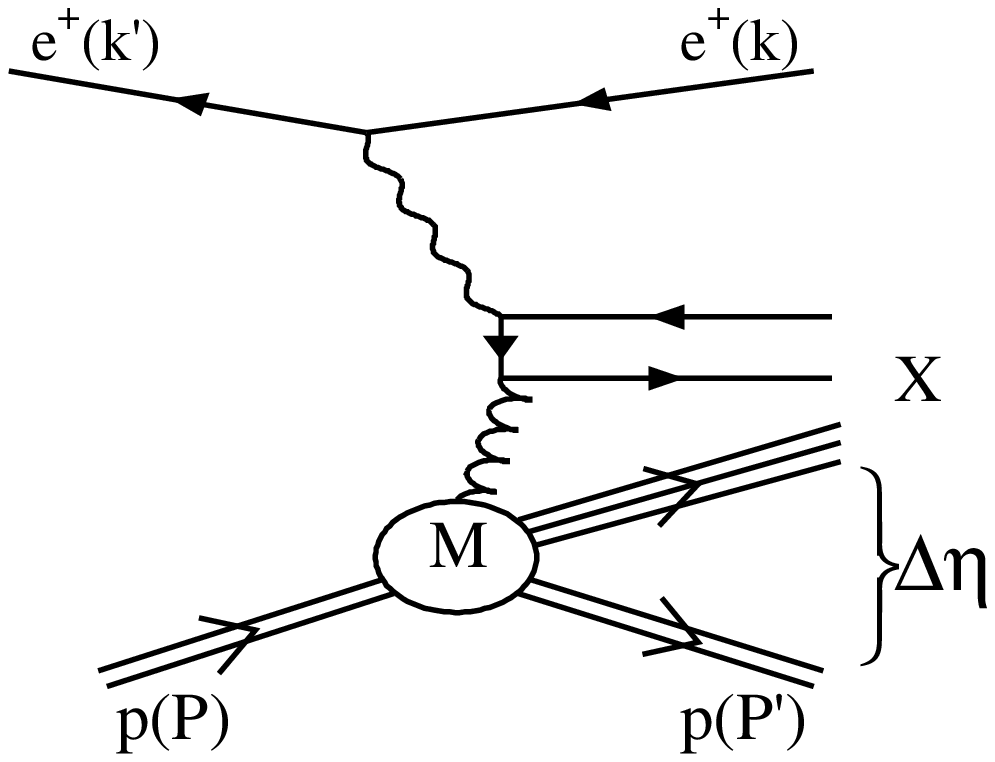}}}
\caption{Direct and Resolved Diffractive Dijet Photoproduction}
\label{fig:photog}
\end{picture}
\end{figure}

Indeed, although hard factorization can be proven
to be valid for diffractive DIS, the proof fails for hadron-hadron
interactions. The posibility of soft interactions taking place
before the hard scattering occurs may spoil factorization in
hadron-hadron collisions\cite{BERR}, although there are no
model-independent estimates of how large the factorization
breaking effects may be or under which circumstances can be found.

Hard factorization has been a very active theoretical topic in the three preceding years, and it is particularly relevant for the discussion of Tevatron diffractive data.

Figure \ref{fig:photo} shows a comparison between H1
dijet photoproduction data and the corresponding prediction computed with
photonic and diffractive parton distributions \cite{photodiff}.
The comparison suggests that for diffractive photoproduction the
factorization breaking mechanisms are either neglible or beyond
the accuracy of the present data.

\setlength{\unitlength}{1.mm} 
\begin{figure}[ht]
\begin{picture}(150,90)(0,0)
\put(40,0){\mbox{\epsfxsize7.0cm\epsffile{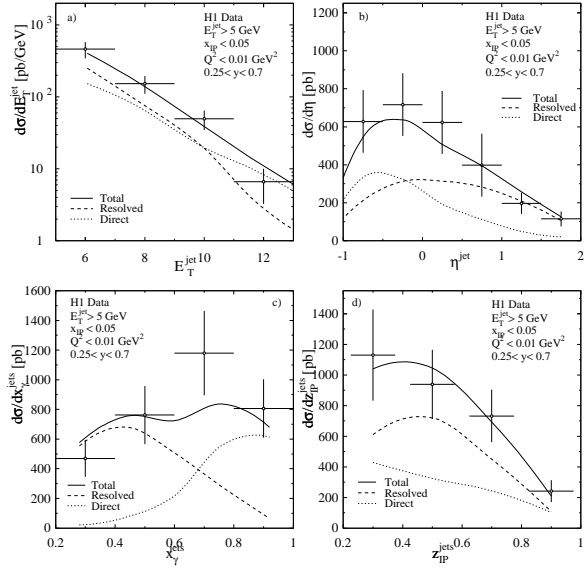}}} \caption{
H1 dijet photoproduction data and the prediction computed with
photonic and diffractive parton distributions} \label{fig:photo}
\end{picture}
\end{figure}

\subsection*{Hard Diffraction at Tevatron}

As we mentioned earlier, large rapidity gaps in proton-antiproton
collisions have been observed by both DO and CDF collaborations at
Tevatron. The typical rapidity gap events observed at Tevatron
are classified acording to three categories: hard single diffraction, hard double pomeron
exchange, and hard colour singlet.

\setlength{\unitlength}{1.mm} 
\begin{figure}[t!]
\begin{picture}(150,50)(0,0)
\put(33,-47){\mbox{\epsfxsize7.5cm\epsffile{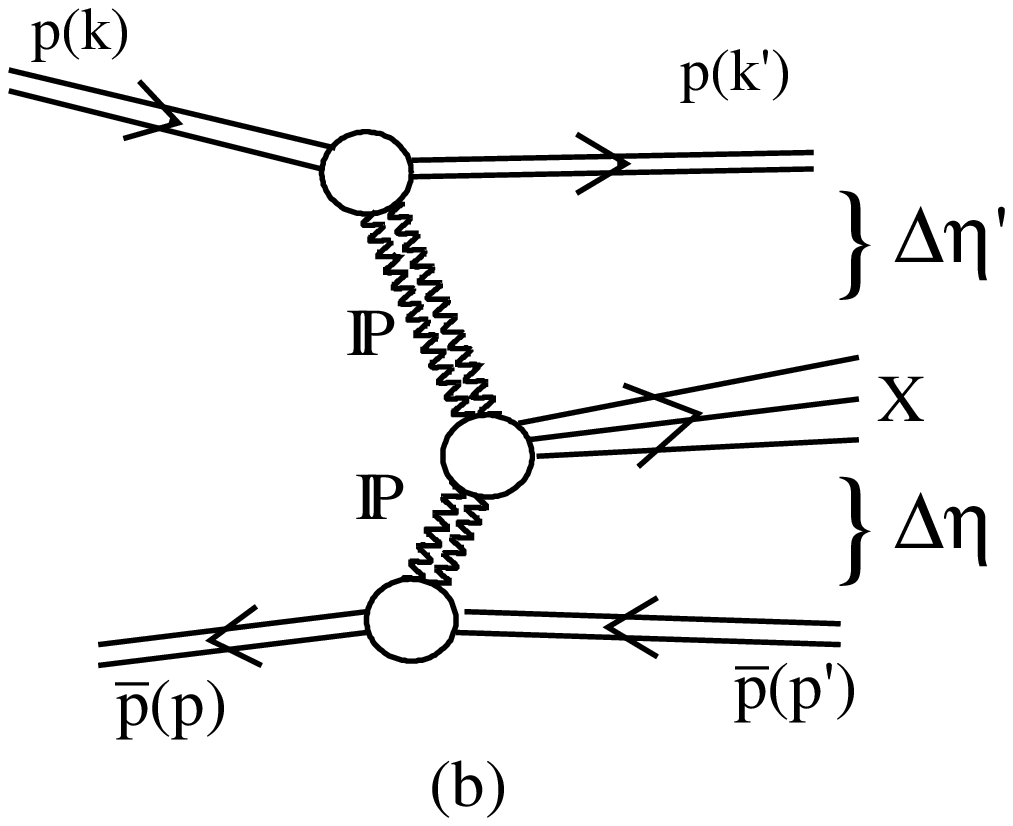}}}
\put(-17,-47){\mbox{\epsfxsize7.5cm\epsffile{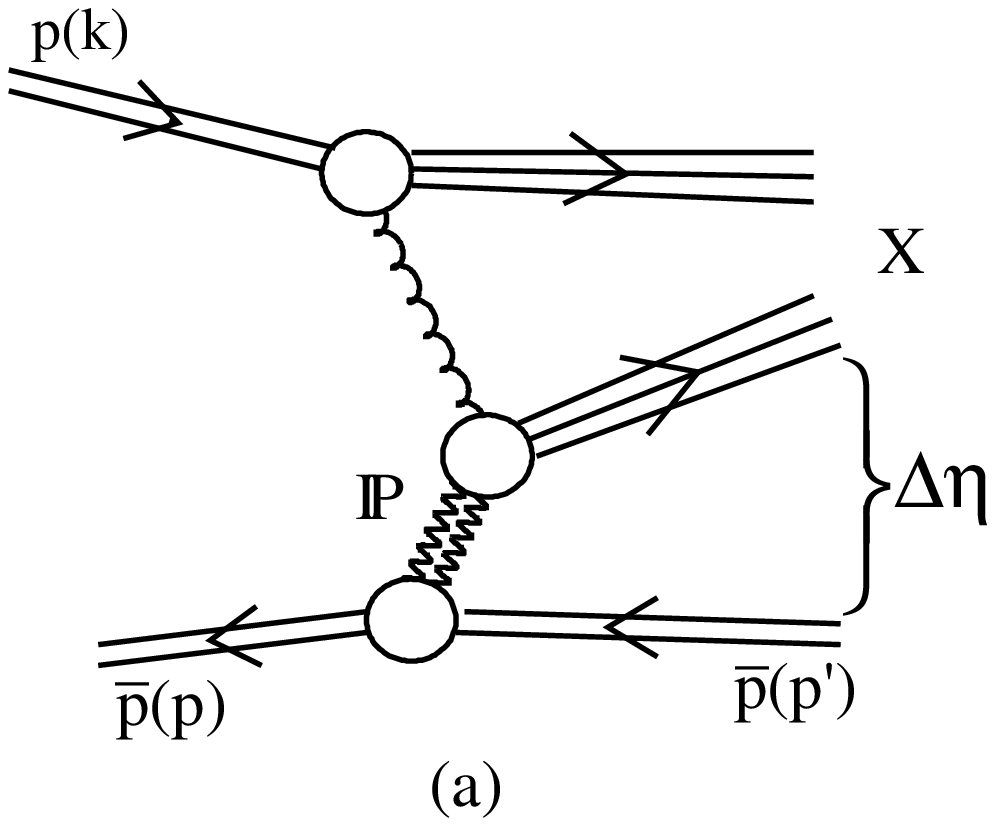}}}
\put(83,-47){\mbox{\epsfxsize7.5cm\epsffile{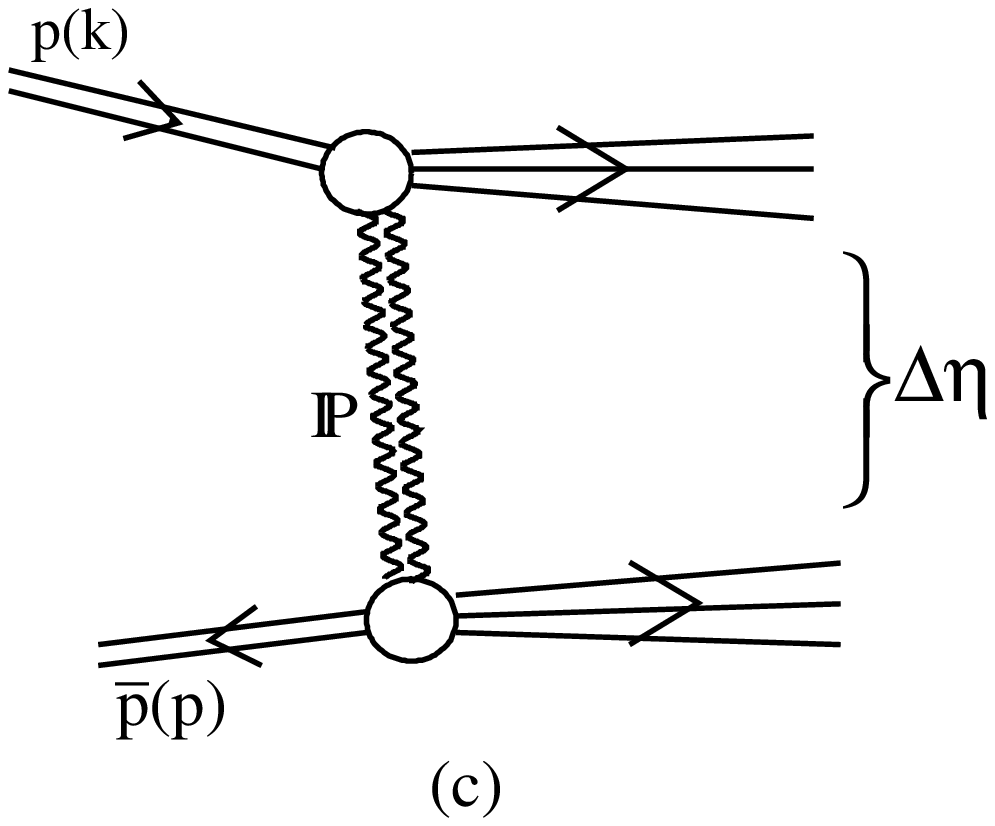}}}
\caption{Rapidity gap topologies at Tevatron.}
\label{fig:TEV}
\end{picture}
\end{figure}

In the first case, Figure \ref{fig:TEV}a, a large rapidity gap in
the forward direction is found between the outgoing antiproton and
the debris produced by the proton-pomeron interaction. In the
double pomeron exchange, Figure \ref{fig:TEV}b, both the proton
and the antiproton survive the interaction and emerge leaving
rapidity gaps between each of them and the debris of the
pomeron-pomeron interaction. In the third case, both the proton
and the antiproton disociate but their corresponding remnants
leave a large rapidity gap between them, Figure \ref{fig:TEV}c.

For a straightforward comparison with HERA data the most simple
topology to analyze is hard single diffraction, as the only change
to be made in the conceptual framework developed in the previous section
is the replacement of the lepton probe (the positron) used at HERA by an hadronic one (the proton) that takes its place at the Tevatron.
The diffractive structure function for
the antiproton is simply related by charge conjugation to that for
the proton, and which is measured at HERA. Whithin this topology, several
observables can be measured such as dijet, $W^{\pm}$, and $Z^0$
production.

Performing this kind of analysis what has been found is that even
though the Tevatron data allow a description in terms of diffractive
parton densities, the predictions computed with those coming from
HERA largely overshot Tevatron data signaling a breakdown of
factorization \cite{Coll}, which remains to be fully understood.

As we have already pointed out in the previous section, the factorization breaking seen Tevatron is not at all unexpected, since QCD factorization
may be broken in hadron-hadron diffraction due to soft exchanges. In addition
to these, there are also some other effects which conspire against factorization
and need some consideration.

For the analysis of these effects it is crucial to
take into account that the standard criterium used to identify diffractive events  both at HERA and at Tevatron, and thus the correspoinding contributions to the diffractive parton distributions and the Tevatron observables, respectively, is the presence of  rapidity gaps of a given size. Obviously,
changing the size of these rapidity gaps, the number of events selected also changes, effect that in fact has been observed and studied by the different
experiments. In Figure \ref{fig:etamax} we show as an example dijet photoproduction events generated with a Monte Carlo in the kinematical regimes of H1 and ZEUS experiments as a function of the pseudo-rapidity $\eta^{max}_{had}$ of the most forward particle belonging to the system $X$ \cite{photodiff}.
Events to the left of the thick vertical line
(rapidity gap limit) are those taken into account while those to the right
are discarted.
\setlength{\unitlength}{1.mm} 
\begin{figure}[t!]
\begin{picture}(100,65)(0,0)
\put(32,-53){\mbox{\epsfxsize9.0cm\epsffile{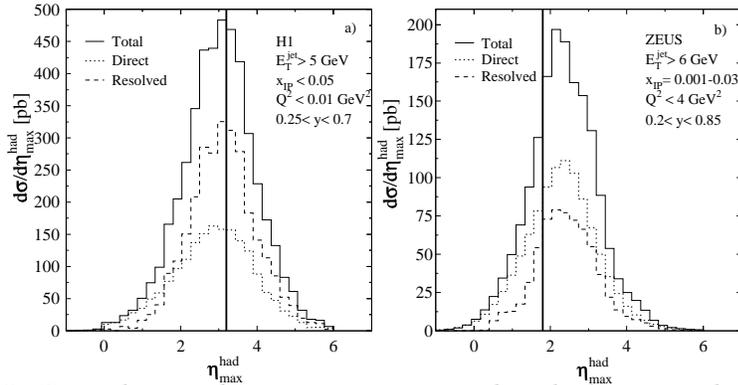}}}
\caption{Dijet photoproduction events generated in the kinematical regimes
of the {\bf a)} H1  and {\bf b)} ZEUS experiments} \label{fig:etamax}
\end{picture}
\end{figure}

It is worthwhile noticing that even for the same kind of process, the
different kinematical regimes covered by both experiments given for
example by cuts in the transverse energy $E_T$ and pseudo-rapidity
$\eta$ of the two most energetic jets, etc., yield rather different
distributions in $\eta^{had}_{max}$ and thus, are affected in a
different way even for the same pseudo-rapidity gap definition. For
ZEUS, approximately 25 \% of the total number of events are
concentrated at $\eta^{had}_{max}< 1.8$, while for H1 kinematics
approximately 10 \% of them survives the same constraint.

Clearly, the situation is much more involved if we try to relate pseudo-rapidity
gap data from different processes as we are doing in the HERA-Tevatron
comparison.,In other words, is not evident what is the relation between
DIS events with a pseudo-rapidity gap of given size, and proton-antiproton events with pseudo-rapidity gaps of a different size, and obtained in a completly different kinematical regime.

It is somewhat paradoxical that the same rapidity gaps that helped so much the development of hard diffraction, seem at this point to be a burden. In any case, ongoing experimental programs based on the Roman Pot technique, which allows
a positive identification of the final state hadron as signature of the colour
singlet exchange, will surely allow a much more deep and complete picture
of these exciting phenomena.

\section*{Connections}

Throughout the present lectures we have review the most significant features
of partons, the fundamental protagonists of QCD, as seen in the three rather
different enviroments, the proton, the photon, and the pomeron (colour singlet exchange). As we have seen, and in fact was anticipated in the introduction, these objects, or more precisely the high energy processes involving them, are three of the main benchmarks of perturbative QCD.

In the process of developing the predictive power of QCD we have introduced
structure, fragmentation, and fracture functions, objects that link intimately both our knowledge and our ignorance about the partonic structure.

The fundamental prediction of QCD in these enviroments, the energy scale dependence of the corresponding cross sections or structure functions, is beautifully illustrated in Figure \ref{fig:Q2} \cite{9812030bb}. The observed behaviour in each case can be traced back to a basic parton interaction mechanism which alternatively dominates over the others: gluon radiation, quark pair creation from gluons, and from photons, respectively, and to the respective parton compositions.

\setlength{\unitlength}{1.mm} 
\begin{figure}
\begin{picture}(150,60)(0,0)
\put(50,10){\mbox{\epsfxsize5.cm\epsffile{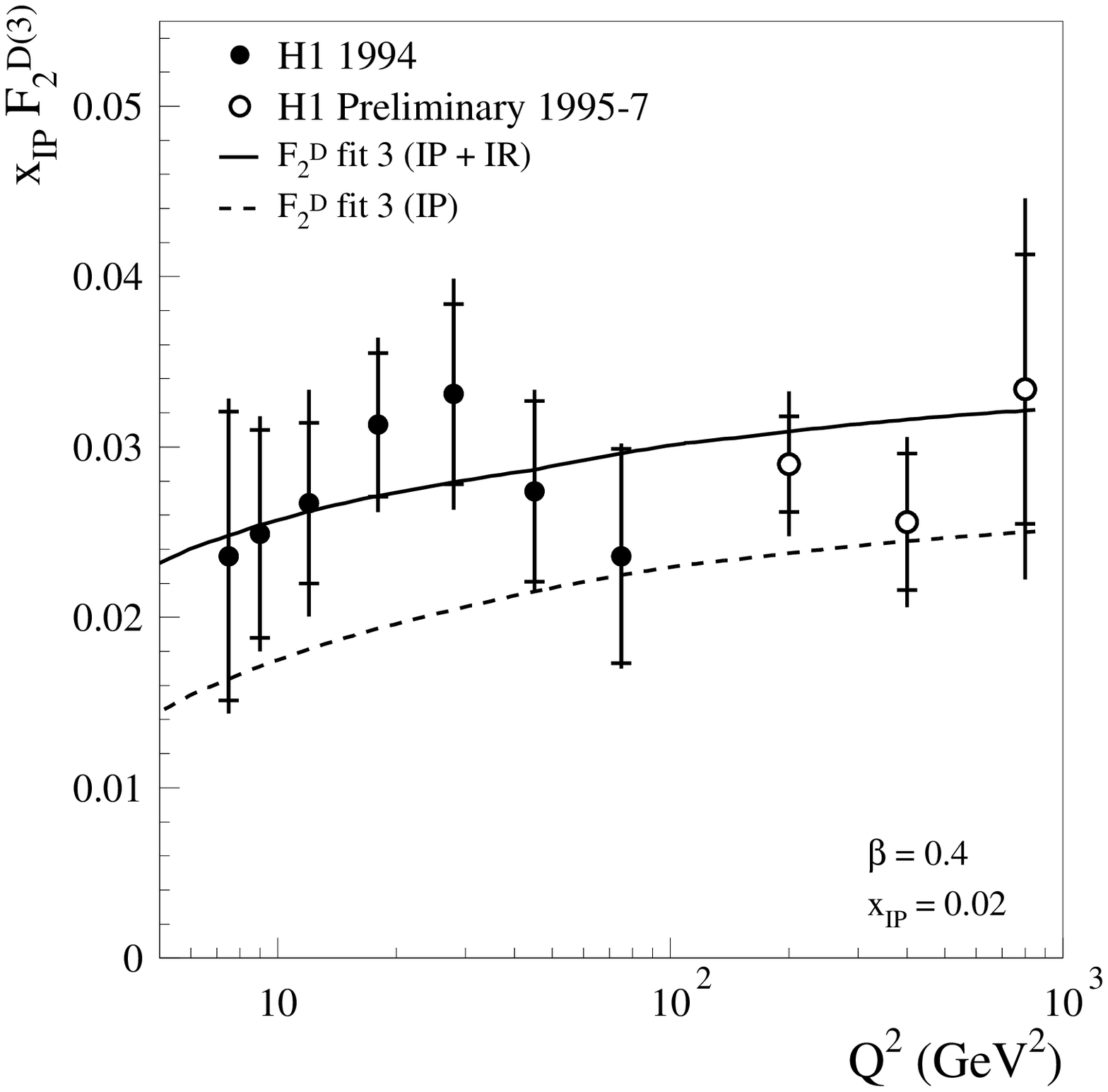}}}
\put(-7,-13){\mbox{\epsfxsize6.5cm\epsffile{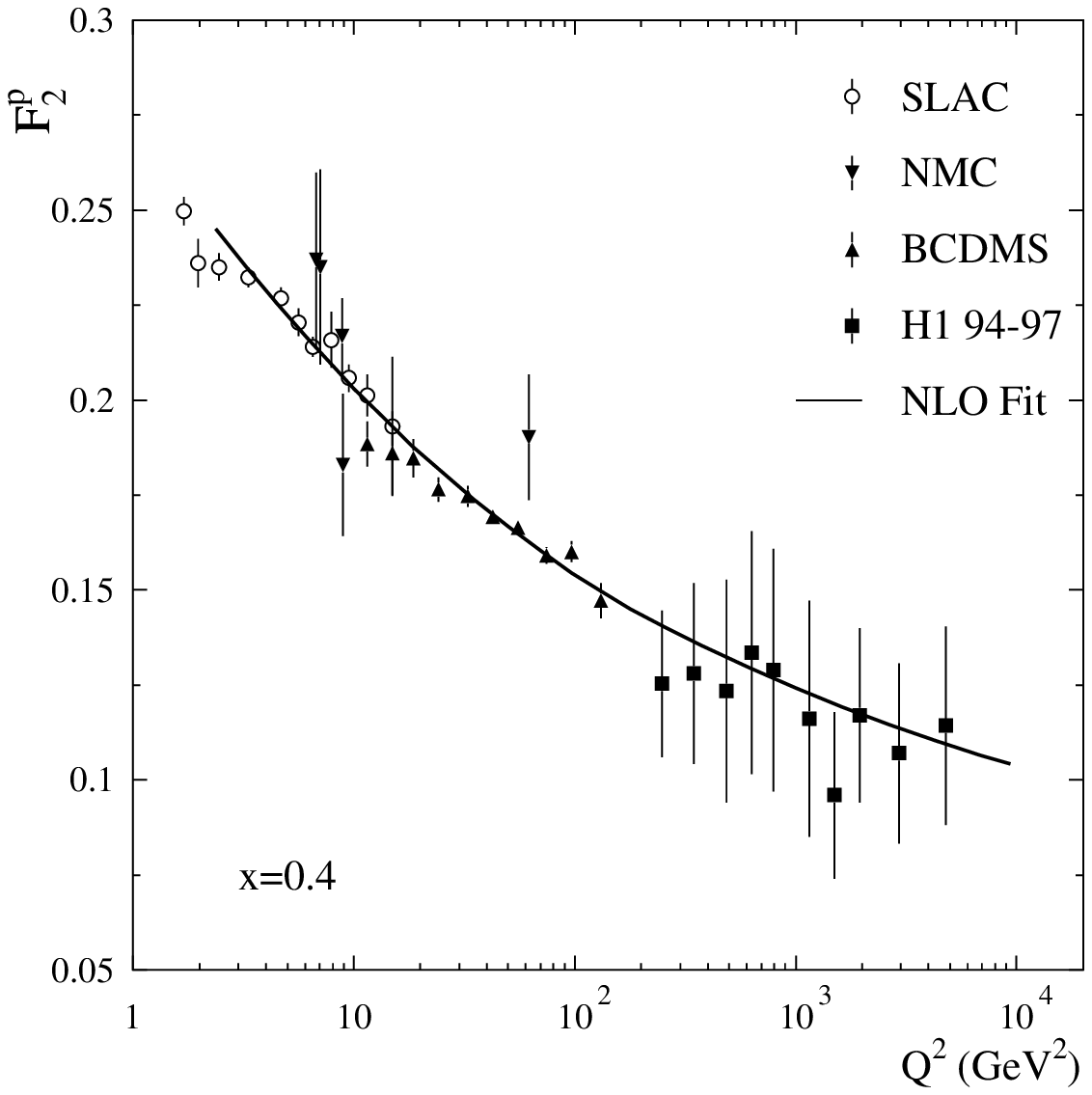}}}
\put(100,9){\mbox{\epsfxsize5.1cm\epsffile{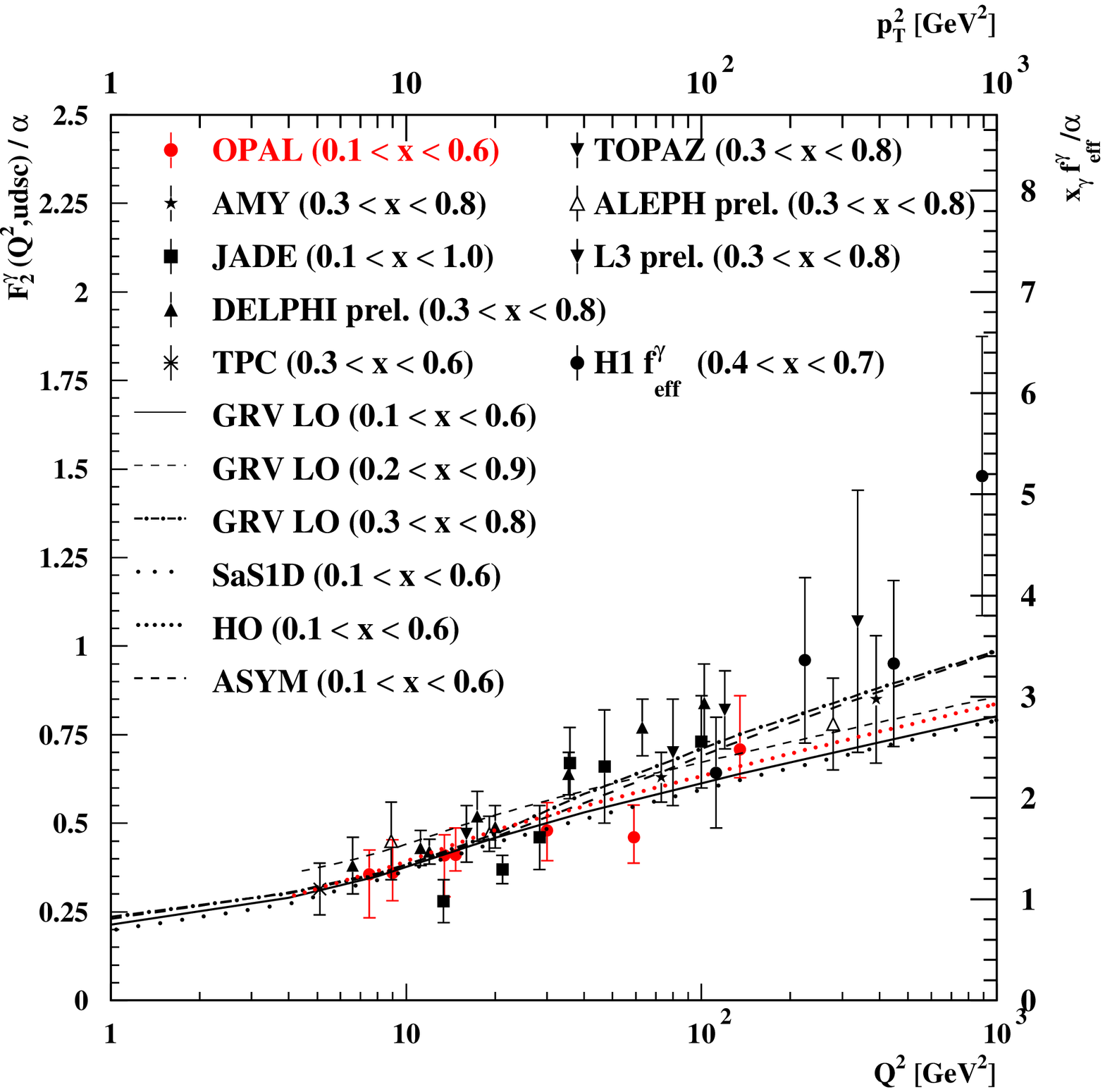}}}
\caption{$Q^2$ Dependence of the proton, the color
singlet, and the photon.} \label{fig:Q2}
\end{picture}
\end{figure}

At intermediate $x$, the case in the figure, the proton structure function is dominated
by valence quarks, however increasing the resolving power of our probe, gluon
radiation depletes dramatically these densities through the corresponding term in the AP evolution equation. At variance with this scenario, the diffractive structure function is dominated by gluons at intermediate $\beta$, and consequently our lepton probes mainly quarks produced by pair creation from gluons, and the corresponding evolution. The logarithmic increase in the photon structure function is caused by
quark pair creation from photons, which in the lowest order approximation appears in the AP equations as $Q^2$ independent term.

Finally, the partonic structure as seen in our benchmarks in some way relate
the physics made in the three main high energy physics laboratories. LEP provides us with the most precise determinations of photon structure functions and also fragmentation functions, using electron-positron collisions. Tevatron in turn, contribute to our understanding of the proton structure and also to diffraction by means proton-antiproton interactions. Positron-proton collisions at Hera, not only test the structure of the proton and diffraction by themselves, but also brings toghether the proton and the photon structure functions in inclusive photoproduction, relates the latter with fracture functions in the diffractive photoproduction, and explores the relation between
fragmentation and fracture functions in leading baryon production.

\end{document}